\documentclass[useAMS,usenatbib]{mn2e}

\usepackage{amssymb}
\usepackage{aas_macros}
\usepackage{natbib}
\usepackage{times}
\usepackage{graphicx}

\newcommand{\Gyr}                      {\,{\rm Gyr}}

\newcommand{\kpc}                      {\,{\rm kpc}}

\newcommand{\Msun}                    {\,{\rm M}_\odot}
\newcommand{\Msunyr}                 {\,{\rm M}_\odot\,{\rm yr}^{-1}}

\newcommand{\hkpc}                     {\,h^{-1}\,{\rm kpc}}
\newcommand{\hMpc}                    {\,h^{-1}\,{\rm Mpc}}
\newcommand{\hMsun}                  {\,h^{-1}\,{\rm M}_\odot}
\newcommand{\kms}                      {\,\,{\rm km}\,\,{\rm s}^{-1}}
\newcommand{\cmcubed}              {\,\,{\rm cm}^{-3}}
\newcommand{\Zsun}                     {\,{\rm Z}_\odot}

\newcommand{\ergs}                     {\,{\rm erg}\,\,{\rm s}^{-1}}
\newcommand{\K}                          {\,{\rm K}}
\newcommand{\keV}                      {\,{\rm keV}}
\newcommand{\nospacehMpc}      {h^{-1}\,{\rm Mpc}}

\newcommand{\gimic}        {\textsc{gimic}}
\newcommand{\owls}        {\textsc{owls}}
\newcommand{\gadget}      {\textsc{Gadget3}}
\newcommand{\subfind}    {\textsc{Subfind}}

\newcommand{\rosat}             {\textit{ROSAT}}
\newcommand{\chandra}         {\textit{Chandra}}
\newcommand{\xmm}             {\textit{XMM-Newton}}

\newcommand{\galexev}         {\textsc{Galaxev}}
\newcommand{\apec}              {\textsc{Apec}}
\newcommand{\ipac}              {\textsc{Ipac}}
\newcommand{\twomass}       {\textsc{2Mass}}

\title[X-ray coronae in simulations of disc galaxy formation]{X-ray coronae in simulations of disc galaxy formation}
 \author[R.~A.~Crain et al.]  {\parbox[h]{160mm} { 
    Robert A. Crain$^{1}$\thanks{E-mail: rcrain@astro.swin.edu.au}, 
    Ian G. McCarthy$^{2}$,
    Carlos S. Frenk$^{3}$,
    Tom Theuns$^{3,4}$ \\ 
    \& Joop Schaye$^{5}$}
  \vspace{6pt}\\
  $^1$Centre for Astrophysics \& Supercomputing, Swinburne
University of Technology, Hawthorn, Victoria 3122, Australia\\ 
  $^2$Kavli Institute for Cosmology, University of Cambridge,
Madingley Road, Cambridge, CB3 0HA\\ 
  $^3$Institute for Computational Cosmology, Department of
  Physics, University of Durham, South Road, Durham, DH1 3LE\\
  $^4$Department of Physics, University of Antwerp, Campus
Groenenborger, Groenenborgerlaan 171, B-2020 Antwerp,  Belgium \\ 
  $^5$Leiden Observatory, Leiden University, PO Box 9513, 2300 RA Leiden, Netherlands}
\begin{document}

\date{\today}
\pagerange{\pageref{firstpage}--\pageref{lastpage}} \pubyear{2009}

\maketitle

\label{firstpage}

\begin{abstract} 
The existence of X-ray luminous gaseous coronae around massive disc galaxies is a long-standing prediction of galaxy formation theory in the cold dark matter cosmogony. This prediction has garnered little observational support, with non-detections commonplace and detections for only a relatively small number of galaxies which are much less luminous than expected. We investigate the coronal properties of a large sample of bright, disc-dominated galaxies extracted from the \gimic\ suite of cosmological hydrodynamic simulations recently presented by Crain et al.  Remarkably, the simulations reproduce the observed scalings of X-ray luminosity with $K$-band luminosity and star formation rate and, when account is taken of the density structure of the halo, with disc rotation velocity as well. Most of the star formation in the simulated galaxies (which have realistic stellar mass fractions) is fuelled by gas cooling from a quasi-hydrostatic hot corona. However, these coronae are more diffuse, and of a lower luminosity, than predicted by the analytic models of White \& Frenk because of a substantial increase in entropy at $z \sim 1-3$. Both the removal of low entropy gas by star formation and energy injection from supernovae contribute to this increase in entropy, but the latter is dominant for halo masses $M_{200} \lesssim 10^{12.5}\Msun$. Only a small fraction of the mass of the hot gas is outflowing as a wind but, because of its high density and metallicity, it contributes disproportionally to the X-ray emission. The bulk of the X-ray emission, however, comes from the diffuse quasi-hydrostatic corona which supplies the fuel for ongoing star formation in discs today. Future deep X-ray observations with high spectral resolution (e.g. with \textit{NeXT/ASTRO-H} or \textit{IXO}) should be able to map the velocity structure of the hot gas and test this fundamental prediction of current galaxy formation theory. 
\end{abstract} 


\section{Introduction}

The rarity of X-ray detections of hot gaseous coronae surrounding disc galaxies poses a fundamental challenge to the current view of how galaxies form. Whilst the possibility that galaxies might contain extra-planar gas was already raised by \citet{Spitzer_56}, it was \citet{White_and_Rees_78} who first proposed that hot gaseous coronae are an integral part of the galaxy formation process.  Their two-stage theory posited that hot gas reservoirs build up as gas condensing onto dark matter haloes is heated by thermodynamic shocks and adiabatic compression. Subsequent radiative cooling of this gas establishes a `cooling flow' that fuels ongoing star formation.

The ubiquity of extended soft X-ray emission from the cooling of galactic gaseous coronae was first predicted by \citet[][hereafter WF91]{White_and_Frenk_91}, whose model built upon the ideas sketched by White \& Rees and established the foundation for interpreting galaxy evolution within the cold dark matter cosmogony. In their framework, the dark matter haloes that host present-day disc-dominated $L^\star$ galaxies engender gas density profiles that are conducive to efficient radiative cooling, via line-emission and thermal Bremsstrahlung, out to radii beyond the optical extent of the central galaxy. The associated cooling rate is sufficient to fuel ongoing star formation in disc galaxies, and thus offset the disruption of discs by mergers \citep{Walker_Mihos_and_Hernquist_96,Barnes_98} and bar-instabilities \citep{Efstathiou_Lake_and_Negroponte_82,Mo_Mao_and_White_98,Syer_Mao_and_Mo_99}, enabling the model to reproduce the abundance of morphological types observed in the local Universe \citep[e.g.][]{Driver_et_al_06_short,Driver_et_al_07,Fukugita_et_al_07_short,Bernardi_et_al_09,Parry_et_al_09}.

The radiation associated with cooling from gas in the haloes of $L^\star$ galaxies is predicted to fall primarily in the soft X-ray band and to have typical surface brightnessess that are readily observable with the \xmm\ and \chandra\ telescopes. This prediction is compelling because analytic and semi-analytic models based upon the WF91 framework successfully reproduce a broad range of galaxy properties, such as the luminosity function in the optical \citep{Cole_et_al_94,Kauffmann_et_al_99,Somerville_and_Primack_99,Cole_et_al_00}, infrared \citep{Lacey_et_al_08} and submillimetre \citep{Baugh_et_al_05} wavebands; the bimodality in the colour-magnitude plane; the apparent `downsizing' of galaxy formation; and the black hole scaling relations \citep{Bower_et_al_06,Croton_et_al_06,De_Lucia_et_al_06}.

The existence of X-ray luminous coronae around massive {\em elliptical} galaxies is well established \citep[e.g.][]{Forman_Jones_and_Tucker_85,Kim_Fabbiano_and_Trinchieri_92,O'Sullivan_Forbes_and_Ponman_01,David_et_al_06,Sun_et_al_07,Jeltema_Binder_and_Mulchaey_08,Sun_et_al_09}. However, this is generally not regarded as a test of the WF91 model because this difffuse X-ray emission could be associated with gas returned to the ISM through stellar evolution \citep[i.e. supernovae and AGB stars, e.g.][]{Read_and_Ponman_98,Mathews_and_Brighenti_98} or with hot gas confined by the potential of galaxy groups - the most common environments of massive ellipticals \citep[e.g.][]{Ponman_et_al_94,Trinchieri_Kim_and_Fabbiano_94,Mulchaey_et_al_96,Trinchieri_Fabbiano_and_Kim_97}. The search for diffuse soft X-ray emission from hot coronae around massive, isolated disc galaxies is therefore the most direct test of the canonical galaxy formation picture.

Adopting this premise, \citet[][hereafter B00]{Benson_et_al_00} analysed \rosat\ observations of three luminous, nearby galaxies, of which two (NGC 2841, NGC 5529) are disc-dominated, whilst the third (NGC 4594, the `Sombrero galaxy') is a bulge-dominated Sa. They found no convincing evidence for diffuse X-ray emission: their upper limits on the soft X-ray luminosity are over an order of magnitude below the luminosity predicted by the WF91 cooling flow model. Similarly, an updated analysis of \chandra\ observations of NGC 5746 by \citet{Rasmussen_et_al_09} failed to produce a significant detection. The greater sensitivity of \xmm\ and \chandra\ relative to \rosat\ has yielded the detection of a diffuse X-ray component for NGC 4594 \citep{Li_et_al_06,Li_Wang_and_Hameed_07}, but even in this case the luminosity shortfall relative to the cooling flow model is approximately two orders of magnitude. Moreover, energy input from supernovae has been claimed as the most likely source of the emission in this case. Detections of diffuse emission from disc-dominated galaxies are, however, now becoming more common \citep[e.g.][]{Strickland_et_al_04,Wang_05,Tullmann_et_al_06,Owen_and_Warwick_09,Sun_et_al_09} and, in general, exhibit a correlation with the star formation rate. A small number of spectacular cases exhibit X-ray emission that has an obvious biconical morphology driven by nuclear starbursts. This has led to the common interpretation that heating by supernovae is the dominant (and perhaps sole) mechanism by which local disc galaxies generate extra-planar X-ray emission, rather than the thermalisation of gravitational potential energy.

Another potential complication is the emission originating from point sources (X-ray binaries in particular), which can contribute a significant fraction of the total X-ray flux.  However, this is much less of a problem for the current generation of X-ray satellites, which have excellent angular resolution (\textit{Chandra} in particular) that enables identification and masking of bright point sources, and high spectral resolution, which allows the removal of the contribution of unresolved, faint point sources \citep[but see][]{Revnitvsev_et_al_08}. Typically, the removal of unresolved point sources is achieved by including a power-law component, in addition to the thermal component, when fitting the X-ray spectrum.

The dearth of convincing X-ray detections of hot coronae is puzzling and suggests that the canonical view of disc galaxy formation is incomplete. Analytic models such as WF91 necessarily rely on simplification of the complex and non-linear processes that are the essential elements of galaxy formation, such as metal-dependent gas cooling, anisotropic gas dynamics, star formation, feedback and chemical enrichment. Potentially, a more general view of the formation of hot coronae and the growth of galaxies is offered by direct, hydrodynamical simulations.

\citet{Toft_et_al_02} highlighted a discord between hydrodynamic simulations and analytic models, finding that efficient radiative cooling can lead to the mass of \textit{hot} gas confined by dark matter haloes being substantially lower than the predictions of cooling flow models.  Since X-ray emissivity is sensitive to gas density ($\propto \rho^2$), the overall X-ray luminosity of the galaxy is commensurately reduced; their model produced galactic coronae with soft X-ray luminosities two orders of magnitude lower than predicted by simple cooling flow models, and consistent with the limited observations available at the time. Interestingly, this result is in contrast to the na\"ive expectation of analytic models where, in general, it is assumed that cooling efficiency and X-ray luminosity are closely correlated.

It is worth noting that \citet{Rasmussen_et_al_09} recomputed the evolution of the systems presented by \citet{Toft_et_al_02}, using an updated simulation code and higher-resolution initial conditions. The more recent study reported X-ray luminosities that were a factor of $\sim 2$ lower than found by \citet{Toft_et_al_02}. Whilst these studies highlight the value of hydrodynamic simulations as a means to test the validity of the simplifications made in analytic models, their differences point to the fact that predictions derived from simulations are also subject to significant uncertainties stemming from resolution effects and the algorithmic implementation of the underlying physics.

In this paper, we investigate the X-ray halo problem using the very large \textit{Galaxies-Intergalactic Medium Interaction Calculation} \citep[\gimic;][hereafter C09]{Crain_et_al_09_short}, a suite of high resolution hydrodynamic resimulations of regions drawn from the Millennium simulation \citep{Springel_et_al_05_short}. A central aim of this project is to study the interaction of galaxies with their gas haloes and with the external intergalactic medium (IGM). The \gimic\ simulations are well suited to our purposes, not only because they feature many well-resolved $L^\star$ galaxies with stellar discs, but also because their convergence behaviour is well understood (see C09), and because the only parameter that was tuned is the mass loading of winds, chosen to ensure a reasonable match to the observed cosmic star formation rate density.

The highest resolution realisations of the simulations represent the stellar discs of galaxies similar to the Milky Way with $\sim 10^5$ particles, whilst the intermediate resolution realisations do so with a factor of 8 fewer. In both cases, the resolution is sufficient to classify the morphology of the galaxies as being either disc- or spheroid-dominated, enabling us to subsample the simulated galaxy population so as to most closely match observational samples. Combined, the intermediate-resolution simulations follow a comoving volume of $1.6\times 10^5\,(\nospacehMpc)^3$, yielding approximately 460 galaxies at $z=0$ with stellar mass greater than $10^{10}\hMsun$. The simulations include mass and energy feedback due to supernovae, and chemical enrichment from type Ia and type II supernovae and asymptotic giant branch (AGB) stars. Radiative cooling is computed on an element-by-element basis under equilibrium conditions in the presence of a realistic photoionising UV/X-ray background.  This allows us to investigate whether X-ray emission is generated primarily by the presence of a gravitationally supported coronae, or by extra-planar gas heated by supernovae. We are also able to assess the importance of metal cooling to the emission.

This work is part of the programme of the Virgo consortium for cosmological simulations. The simulations adopt the same cosmological parameters as the Millennium simulation \citep{Springel_et_al_05_short}: $\Omega_{\rm m} = 0.25$, $\Omega_\Lambda = 0.75$, $\Omega_{\rm b} = 0.045$, $n_{\rm s}=1$, $\sigma_8 = 0.9$, $H_0 = 100~h~{\rm km~s}^{-1}~{\rm Mpc}^{-1}$, $h=0.73$. Our choice of cosmological parameters is consistent with the most recent determinations from the cosmic microwave background radiation and other cosmological tests, with the exception of $\sigma_8$, which is higher by $\sim2\sigma$. We do not expect this difference to affect the results and conclusions of this work, since for the halo mass range considered here it results in only a small change in the epoch of formation. 

The paper is laid out as follows. In \S~\ref{sec:methods} we describe our methods, giving a brief overview of the \gimic\ simulations and our main postprocessing techniques. In \S~\ref{sec:comp_with_obs} we confront the simulations with observational constraints, whilst in \S~\ref{sec:theory} we explore the analytic theory and compare it with the results of the hydrodynamic simulations. We summarise and discuss the results in \S~\ref{sec:summary}. A short appendix is included, in which we discuss the sensitivity of our numerical results to resolution. 


\section{Methods}
\label{sec:methods}

\subsection{Simulations}
\label{sec:simulations}

The \gimic\ simulations are described in detail in C09, where thorough discussions of the generation of the initial conditions, the simulation code, and the initial results may be found. Here, we present only a brief overview and limit the description to aspects that are specifically relevant to this study.

\gimic\ is designed to circumvent the unfeasibility - due to computational expense - of simulating large cosmological volumes ($L \gtrsim 100\hMpc$) at high resolution ($m_{\rm gas} \lesssim 10^7\hMsun$) to $z=0$. Using `zoomed' initial conditions \citep{Frenk_et_al_96,Power_et_al_03,Navarro_et_al_04_short}, \gimic\ follows, with full gas dynamics, the evolution of five roughly spherical regions drawn from the Millennium simulation. In order to encompass a wide range of large-scale environments, the regions were chosen such that their overdensities deviate by $(-2$, $-1$, $0$, $+1$,$+2)\sigma$ from the cosmic mean, where $\sigma$ is the rms mass fluctuation, on a scale of $18\hMpc$, at $z=1.5$. The $+2\sigma$ region was additionally constrained by the requirement that it be centred on a rich galaxy cluster halo. In practice, this ensures that the simulations include rare cosmological features, since the $-2\sigma$ region is also approximately centred on a sparse void. Each region has an approximate comoving radius of $18\hMpc$, except the $+2\sigma$ region which was enlarged to a radius of $25\hMpc$ in order to accommodate the rich cluster. The remainder of the $500^3\,(\nospacehMpc)^3$ Millennium simulation volume is modelled with collisionless particles at much lower resolution in order to follow the large-scale structure. The scheme is illustrated schematically by Fig.~1 of C09.

The initial conditions were realised at intermediate resolution ($m_{\rm gas} = 1.16\times 10^7\hMsun$) and high resolution ($m_{\rm gas} = 1.45\times 10^6\hMsun$); we reserve the term `low resolution' for the original Millennium simulation, in which the collissionless particles, representing a composite of baryonic and dark matter, have mass $8.6\times10^8\hMsun$. Gravitational forces on the baryonic and high-resolution dark matter particles were softened over an identical length scale; we adopt a softening length that is initially fixed in comoving coordinates, but becomes fixed in physical coordinates at a predefined redshift, i.e. $\epsilon_{\rm com}(a)^{'} = \min(\epsilon_{\rm com},\epsilon_{\rm phys}^{\rm max}/a)$. The softenings were chosen such that at $z=3$, they are fixed at $\epsilon_{\rm phys}^{\rm max} = (1.0,0.5)\hkpc$ for the intermediate- and high-resolution runs, respectively.

The simulations were performed with the TreePM-SPH code \gadget, a substantial upgrade of \textsc{Gadget2} \citep{Springel_05} that includes:

\begin{enumerate} \item a new domain decomposition algorithm that improves load balancing, particularly for simulations with strongly clustered particle distributions run on parallel supercomputers with a large number of cores \citep{Springel_et_al_08_short}; \item a prescription for star formation designed to enforce a local Kennicutt-Schmidt law \citep{Schaye_and_Dalla_Vecchia_08}; \item the contribution of metals to the cooling of gas, computed element-by-element, in the presence of an imposed UV-background \citep{Wiersma_Schaye_and_Smith_09}; \item stellar evolution and the associated delayed release of 11 chemical elements \citep{Wiersma_et_al_09}; \item galactic winds that pollute the IGM with metals and can quench star formation in low-mass haloes \citep{Dalla_Vecchia_and_Schaye_08}.  \end{enumerate}

The code does not, however, model the evolution of black holes or feedback effects associated with them. The hydrodynamics implementation, taken from \textsc{Gadget2}, is the entropy conserving formulation of smoothed particle hydrodynamics \citep[SPH;][]{Gingold_and_Monaghan_77,Lucy_77}, as discussed in \citet{Springel_and_Hernquist_2002}. For further details, the reader is referred to C09 and references therein. All five regions were evolved to $z=0$ at intermediate resolution. Only the $-2\sigma$ region was evolved to $z=0$ at high resolution (the $-1\sigma$, $0\sigma$ and $+1\sigma$ regions were run to $z=2$; the $+2\sigma$ region was not run at high resolution). Since we are interested in the low redshift evolution of galaxies, we follow the philosophy of C09 and consider the five intermediate-resolution runs as a fiducial simulation set, and use the high-resolution realisation of the $-2\sigma$ simulation to assess how sensitive our results are to numerical resolution. An assessment of numerical convergence is supplied in the Appendix.

Owing to its unique initial conditions, \gimic\ represents a complementary simulation suite to the Overwhelmingly Large Simulations \citep[\owls;][]{Schaye_et_al_10_short}. Both projects are based on the same simulation code. The aim of \owls\ is to investigate the dependence of various properties on the parametrisation of subgrid physics, whilst \gimic\ accesses a wider range of environments for a single physics implementation of \owls\ (the \textsc{mill} simulation in \owls). 

We note that we are concerned here with the establishment of hot coronae, for which shock heating (induced by gravitational collapse and the impact of galactic winds with the intergalactic medium) can play a key role. The ability of our hydrodynamics scheme to model this process robustly is therefore important. SPH has been shown to reproduce the analytic solutions of simple problems such as shock tubes, spherical collapse, and Sedov blasts \citep[e.g.][]{Springel_05,Tasker_et_al_08}. Tests of the formation of cosmological haloes assuming non-radiative hydrodynamics have demonstrated that SPH produces similar results to those of Eulerian adaptive mesh refinement (AMR) schemes beyond the central 10~percent of the halo virial radius \citep{Frenk_et_al_99_short,Kravtsov_Nagai_and_Vikhlinin_05,Voit_Kay_and_Bryan_05}. This agreement in the limit of non-radiative physics therefore suggests that, in spite of its inability to resolve shocks well, SPH schemes model gravitational shock heating adequately.


\subsection{Halo and galaxy identification}

As decribed in C09, we identify bound haloes using the \subfind\ algorithm presented by \citet{Dolag_et_al_08}, which extends the standard implementation \citep{Springel_et_al_01} by also considering baryonic particles when identifying self-bound substructures. This procedure first finds dark matter haloes using a friends-of-friends (FoF) algorithm, with the standard linking length in units of the interparticle separation \citep[$b=0.2$;][]{DEFW85}. It also associates all baryonic particles with their nearest neighbour dark matter particle. The aggregated properties of baryonic particles associated with grouped dark matter particles define the baryonic properties of each FoF group. Halo substructures are then identified using a topological unbinding algorithm. This provides an unambiguous definition of a \textit{galaxy} within the simulations, namely the set of star particles bound to individual subhaloes. The gas bound to each subhalo then forms the interstellar medium (ISM) and the hot corona. Since individual haloes may have more than one `subhalo', they may host more than one galaxy. In analogy to semi-analytic models, the stars associated with the most massive subhalo of a FoF group are hence defined as central galaxies, whilst stars associated with substructures are satellites.


\subsection{Morphological classification}
\label{sec:morphological_classification}

The resolution of our simulations is sufficient to allow a morphological classification of the galaxies into disc- and spheroid-dominated types, based on their dynamics. We assume a simple two-component model: i) a dispersion-supported spheroid, and ii) a rotationally-supported disc. The centre of (baryonic + dark) mass of the main subhalo of each FoF group is used as the starting point for an iterative procedure that computes the centre of mass of all star particles within a sphere that, at each iteration, is decreased in radius by 2 per cent and re-centred on the new centre of stellar mass. The procedure concludes when the sphere encloses fewer than 32 star particles; we have checked that this scheme yields robust galactic centres. The smaller of three times the stellar half mass radius, $r_{1/2}$, or the distance to the furthest bound star particle is then used as a `boundary' for the galaxy, $r_{\rm gal}$. This is to exclude the contribution of the diffuse `intrahalo stars' that \subfind\ associates with the potential of the most massive halo substructure. This is an important practice, since our sample includes a number of massive galaxies for which up to 30 per cent of the bound stellar mass is (by this definition) in the form of intrahalo stars.  

\begin{figure}
\includegraphics[width=\columnwidth]{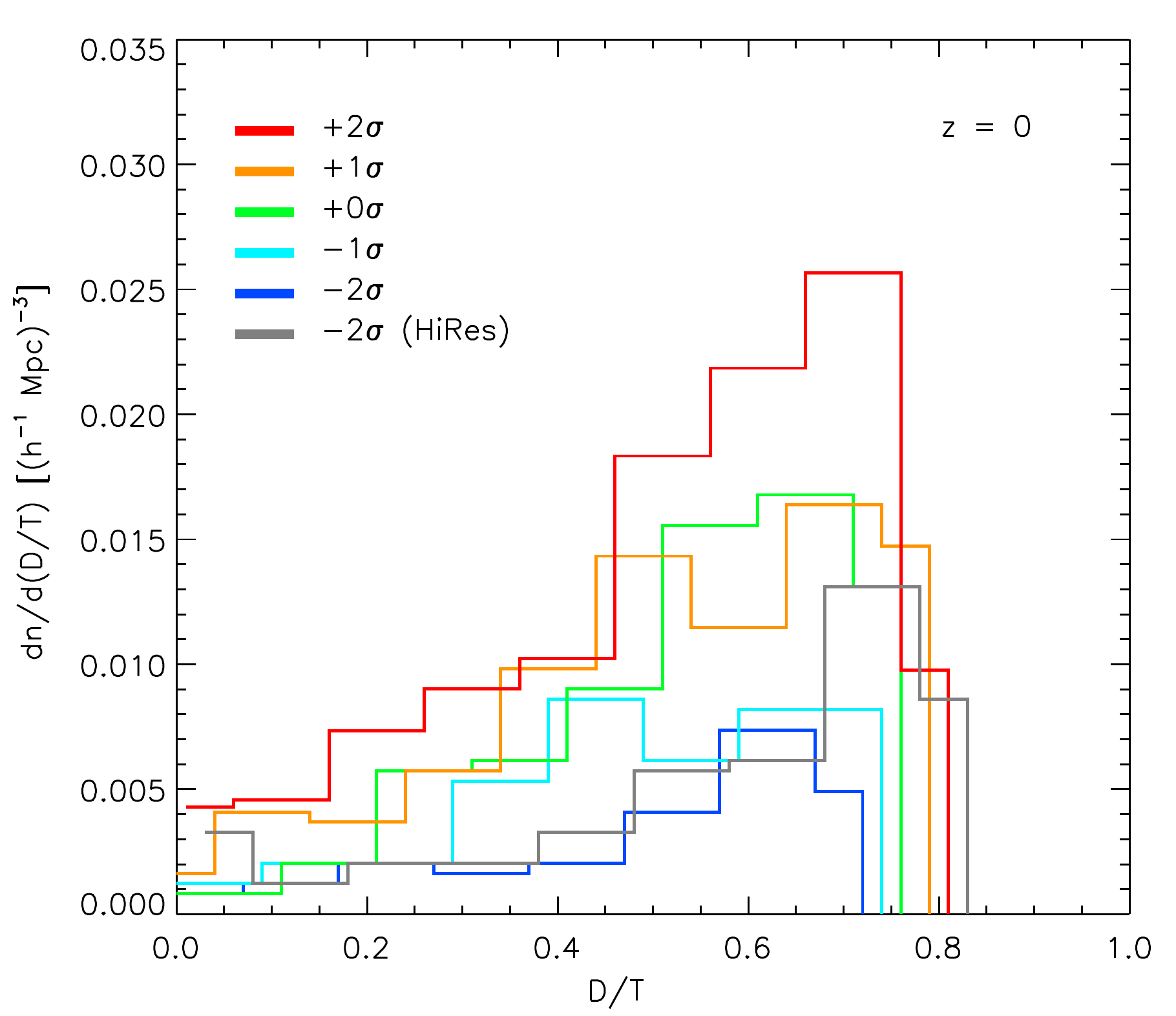}
\caption{Histogram of disc-to-total stellar mass ratios (D/T) for the five \gimic\ regions. Results from the high-resolution $-2\sigma$ region are also shown to illustrate the sensitivity of morphology to resolution. Approximately half of all galaxies in the simulations are disc-dominated (i.e. D/T $> 0.5$) at $z=0$. To create a sample that mimics ``late-type'' galaxies we adopt a criterion of D/T $> 0.3$ for inclusion in our sample.} 
\label{fig:DT_Histogram}
\end{figure}

To decompose the stellar mass of galaxies into spheroid and disc components, we use a procedure based on the method introduced by \citet{Abadi_03}. We compute the angular momentum, $\mathbf{L}$, of the $N_\star$ star particles within $r_{\rm gal}$: \begin{equation} \mathbf{L} = \sum_{i=1}^{N_\star}\mathbf{r}_i \times \mathbf{p}_i, \label{eq:ang_mom} \end{equation} where $\mathbf{r}$ is the radial vector with respect to the centre of the system and $\mathbf{p}$ is the linear momentum vector corrected for the bulk peculiar velocity of the system. The assumption that the spheroid is fully dispersion supported requires that it should have no net angular momentum, and so the spheroid mass can be reasonably approximated as twice the summed mass of particles that counter-rotate with respect to $\mathbf{L}$. The remaining stellar mass of the galaxy then comprises the disc. 

The distribution of D/T for all galaxies with $M_\star > 10^{10}\Msun$ is shown, for the five \gimic\ regions, in Fig.~\ref{fig:DT_Histogram}. The total number of galaxies included from the five intermediate resolution regions is 1267, whilst 111 galaxies are included from the high-resolution $-2\sigma$ region. In each region, approximately half of all galaxies are disc-dominated (D/T $>0.5$); this result is consistent with the morphological analyses of the Millennium Galaxy Catalogue by \cite{Driver_et_al_06_short} and of the SDSS catalogue by \cite{Benson_07}. Since we wish to create a sample that excludes only obviously ``elliptical'' galaxies, we adopt a slightly lower threshold and define ``disc galaxies'' as those with a disc-to-total stellar mass fraction $>0.3$. The precise choice of this fraction is unimportant for the purposes of this study; we have checked that our results are not affected by adopting a value of 0.2 or 0.5 instead.

The high proportion of disc-dominated galaxies is a noteworthy success of our simulations. It is a well-known problem of hydrodynamical simulations of this type that excessive cooling in small halos, allied to angular momentum transport during mergers, results in discs with much lower angular momentum than observed for real galaxies \citep{Weil_Eke_and_Efstathiou_98}. Recently, high-resolution simulations of individual objects have demonstrated the importance of feedback in mitigating this problem \citep[e.g.][]{Okamoto_et_al_05,Scannapieco_et_al_08}, a finding supported by the large sample of galaxies we present here. We plan to investigate this interesting result in future work.


\subsection{Computing luminosities}

\begin{figure} 
\includegraphics[width=\columnwidth]{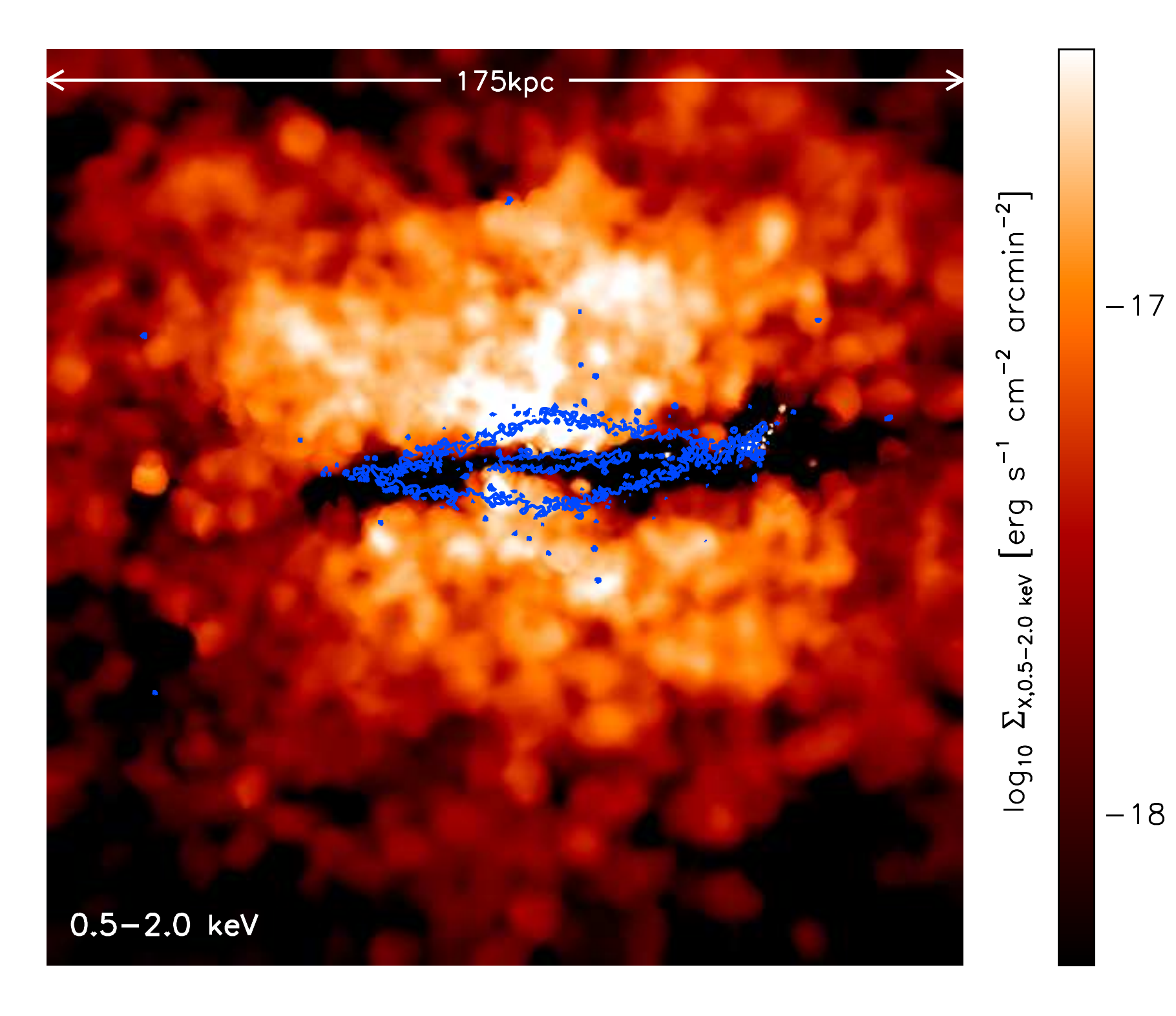} 
\caption{Synthetic surface brightness map at $z=0$ of a galaxy drawn from our sample of \gimic\ galaxies. The map is $175\kpc$ in each dimension, projected over the same depth. The galaxy is edge-on to the projection axis. The colour mapping shows the surface brightness of soft X-ray emission, whilst the blue contours trace the underlying $K$-band surface brightness of the galaxy.} 
\label{fig:map} 
\end{figure}

Optical and X-ray luminosities are computed in postprocessing. The former are calculated individually for star particles, considering them as simple stellar populations (SSP). The masses of individual stellar particles produced in our simulations ($\sim10^6-10^7\Msun$) are comparable to, or greater than, the typical mass of star clusters, so it is reasonable to assume that they can be described as an SSP. We assume a universal initial mass function \citep[IMF;][]{Chabrier_03} and store the age and metallicity of star particles. Thus a spectral energy distribution (SED) for each particle can be derived by interpolation over the \galexev\ models of \citet{BC03}. The optical luminosity of each star particle is obtained by integrating the product of its SED with the appropriate filter transmission function; the overall broadband luminosity of a galaxy is then defined as the sum of the luminosities of all star particles within $r_{\rm gal}$. Since we are primarily concerned here with $K$-band luminosities, we do not expect dust extinction to be important.

Gas phase X-ray luminosities are also computed on a per-particle
basis.  We arrive at the overall X-ray luminosity of a galaxy by summing the luminosities of all gas particles bound to its subhalo. The X-ray luminosity of the $j^{\rm th}$ gas  particle is computed as: 
\begin{eqnarray}
  L_{\rm X,j} & = & n_{{\rm e},j} n_{{\rm H},j} \Lambda_j V_j\\
           & = & \frac{X_{\rm e}(Z_j)}{[X_{\rm e}(Z_j)+X_{\rm i}(Z_j)]^2}
\biggl(\frac{\rho_j}{\mu(Z_j) m_{\rm H}} \biggr)^2 \Lambda_j V_j \nonumber \\ 
           & = & \frac{X_{\rm e}(Z_j)}{[X_{\rm e}(Z_j)+X_{\rm i}(Z_j)]^2} \frac{\rho_j}{\mu(Z_j) m_{\rm H}} 
                 \frac{m_{{\rm gas},j}}{\mu(Z_j) m_{\rm H}} \Lambda_j, \nonumber
\end{eqnarray}
\noindent where $\rho$ is the particle gas density, $m_{\rm gas}$ is the particle mass, the volume is $V = m_{\rm gas} / \rho$, $n_{\rm e}$, $n_{\rm H}$ and $n_{\rm i}$ are the number densities of electrons, hydrogen, and ions, respectively, $X_{\rm e} \equiv n_{\rm e}/n_{\rm H}$, $X_{\rm i} \equiv n_{\rm i}/n_{\rm H}$, $Z$ is the metallicity, $\mu$ is the mean molecular weight, $m_{\rm H}$ is the mass of a hydrogen atom, and $\Lambda$ is the cooling function in units of ergs cm$^3$ s$^{-1}$ (integrated over some appropriate passband, such as 0.5-2.0\keV).  We compute $\Lambda$ by interpolating a pre-computed table generated using the Astrophysical Plasma Emission Code\footnote{ To maintain strict consistency with the implementation of radiative cooling in the simulations, it would be more appropriate to use cooling rates predicted by the CLOUDY software package (Ferland et al.\ 1998).  The \apec, however, is more widely used in the analysis of X-ray data, which is why we have adopted it here. We verified that using CLOUDY instead gives nearly identical results.} \citep[\apec, v1.3.1, see][]{Smith_et_al_01} under the assumption that the gas is an optically thin plasma in collisional ionisation equilibrium.  \apec\ cooling rates are computed on an element-by-element basis and summed to yield the total cooling rate of each particle, i.e.,
\begin{equation}
\Lambda_j(T_j) = \sum_{k=1}^{N} \lambda_{j,k}(T_j), 
\end{equation}
\noindent where $T$ is the gas temperature and $\lambda_{j,k}(T_j)$ is the cooling function for element species $k$ for the $j^{\rm th}$ particle. The summation is performed over the 11 most important elements for cooling (H, He, C, Ca, N, O, Ne, Mg, S, Si, Fe), which are individually and self-consistently tracked during the simulation. \apec\ assumes the solar abundance ratios of \citet{Anders_and_Grevesse_89}, but it is straightforward to modify the spectra for arbitrary abundances, which we have done. Note that our scheme automatically excludes gas within the cool disc of the interstellar medium (ISM, i.e. those for which $n_{\rm H} > 0.1\cmcubed$) since we assign these particles a temperature of $T=10^4\K$, which is below the minimum temperature for which \apec\ returns a non-zero X-ray luminosity. Therefore we consider, by construction, only extra-planar emission.

It should be noted that the \apec\ cooling tables assume pure collisional ionisation equilibrium and hence neglect the extragalactic UV/X-ray background. However, the effect of photoheating {\it on the derived X-ray properties} is small in the regime we explore here.  \citet{Wiersma_Schaye_and_Smith_09} show that the difference in cooling time for hot ($T > 2\times 10^5\K$) plasmas under collisional and photoionisational equilibrium, with and without photoheating, is essentially negligible at all densities in the absence of metals, but for low-density plasmas ($n_{\rm H} \lesssim 10^{-4}\cmcubed$) enriched to solar abundance, the cooling time can be a factor of $\gtrsim 2$ greater in the presence of a metagalactic photoionising UV/X-ray flux. However, we find that only a small fraction of the total X-ray luminosity of $L^\star$ haloes comes from gas at such low densities. It is important to note though that this does not imply that photoheating is unimportant in the formation and evolution of disc galaxies.  Indeed, it is possible that the progenitors of these systems could have been substantially affected by the UV/X-ray background.


\subsection{The sample of simulated galaxies}
\label{sec:sample}

\begin{figure}
\includegraphics[width=\columnwidth]{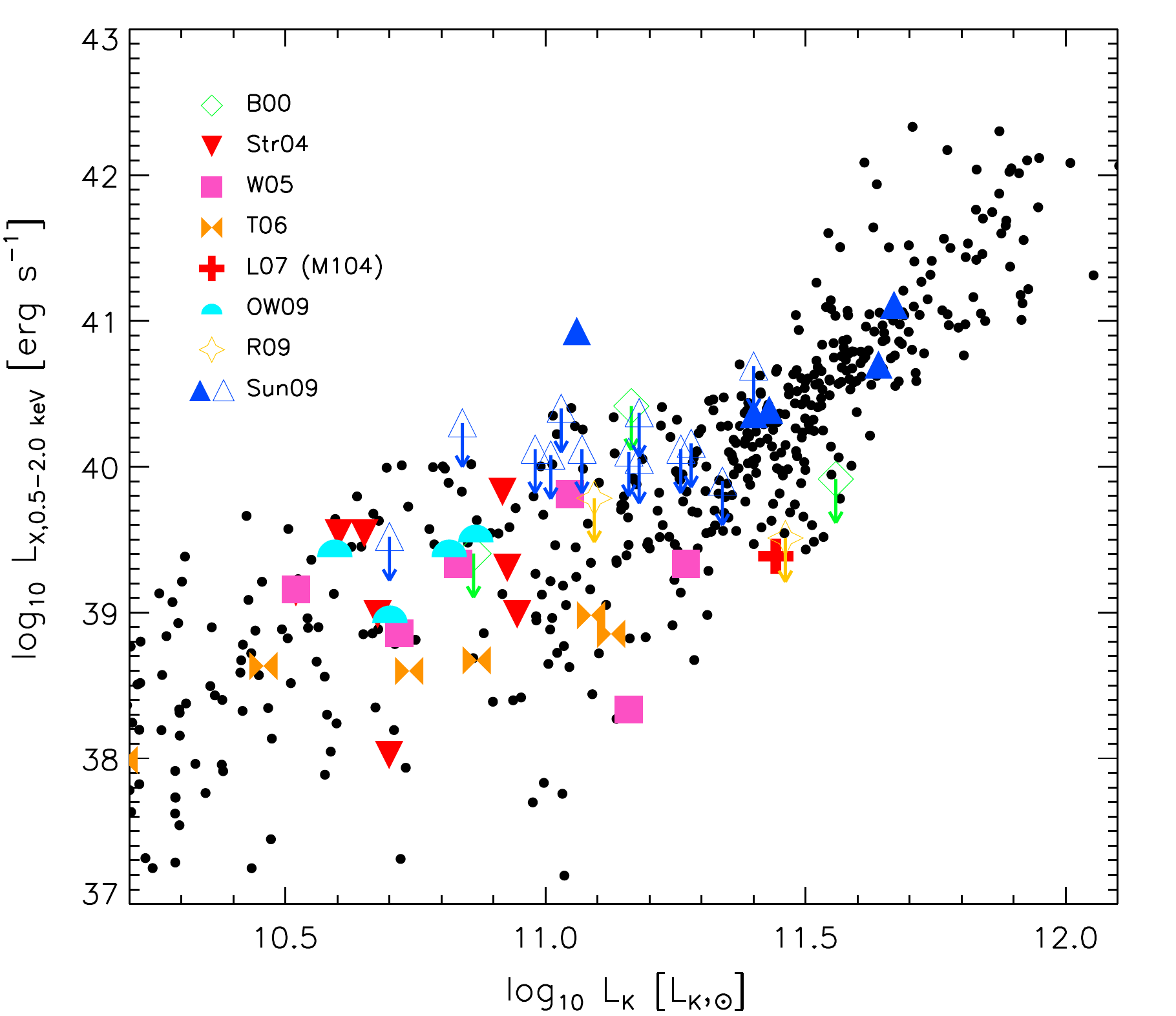}
\caption{The present-day 0.5-2.0 keV X-ray luminosity as a function of $K$-band luminosity. Simulated galaxies are represented by black dots. For the observational data, extra-planar point source-corrected X-ray luminosities have been extracted from a number of recent studies (see legend and text).  The corresponding $K$-band luminosities of these galaxies were extracted from the online \twomass\ database at \ipac. The simulations reproduce both the scaling and the scatter of the observations.} 
\label{fig:Lx_Lk}
\end{figure}

In the following section, we compare the results of the \gimic\ simulations with observational measurements. The galaxies are drawn from all 5 intermediate-resolution \gimic\ regions, and we impose a number of selection criteria in order to construct a sample similar to observational samples. We select central (i.e. the most massive galaxy within a FoF halo), disc galaxies (D/T $> 0.3$) whose stellar mass lies in the range $10^{10} < M_\star < 10^{11.7}\Msun$, roughly corresponding to $L^\star$ at $z=0$. Additionally, we require that the subhalo with which each galaxy is associated should account for at least $90~$per cent of the mass of the parent FoF group, i.e. $M_{\rm sub} > 0.9M_{\rm FoF}$. This criterion is intended to select only isolated systems by excluding galaxies that are interacting or are members of galaxy groups and clusters. The 5 regions yield a sample of 458 galaxies fulfilling these criteria. 

To illustrate the correspondence of our methods with observational techniques, we show in Fig.~\ref{fig:map} a synthetic surface brightness map of the diffuse soft X-ray and $K$-band optical emission, computed for a galaxy drawn from our sample. The galaxy has the following properties: $M_{200} = 10^{12.0}\Msun$, $M_\star = 10^{10.6}\Msun$, $L_{\rm K} = 10^{10.8}{\rm L}_{{\rm K},\odot}$, $L_{\rm X,0.5-2.0 keV}=10^{39.1}{\rm erg~s}^{-1}$. In general the surface brightness increases towards the galactic centre, in keeping with the WF91 picture, but close to the galaxy disc the X-ray emission structure becomes more complex, highlighting the necessity of employing hydrodynamical simulations to calculate the detailed evolution of the gas. Extra-planar emission is clearly visible, with the contours tracing the dense disc of the ISM at small radii and becoming more circular (in projection) at larger radii, except where their morphology is disturbed by accreting substructure.


\section{X-ray coronae of disc galaxies: comparison with observations}
\label{sec:comp_with_obs}

\begin{figure*}
\includegraphics[width=\textwidth]{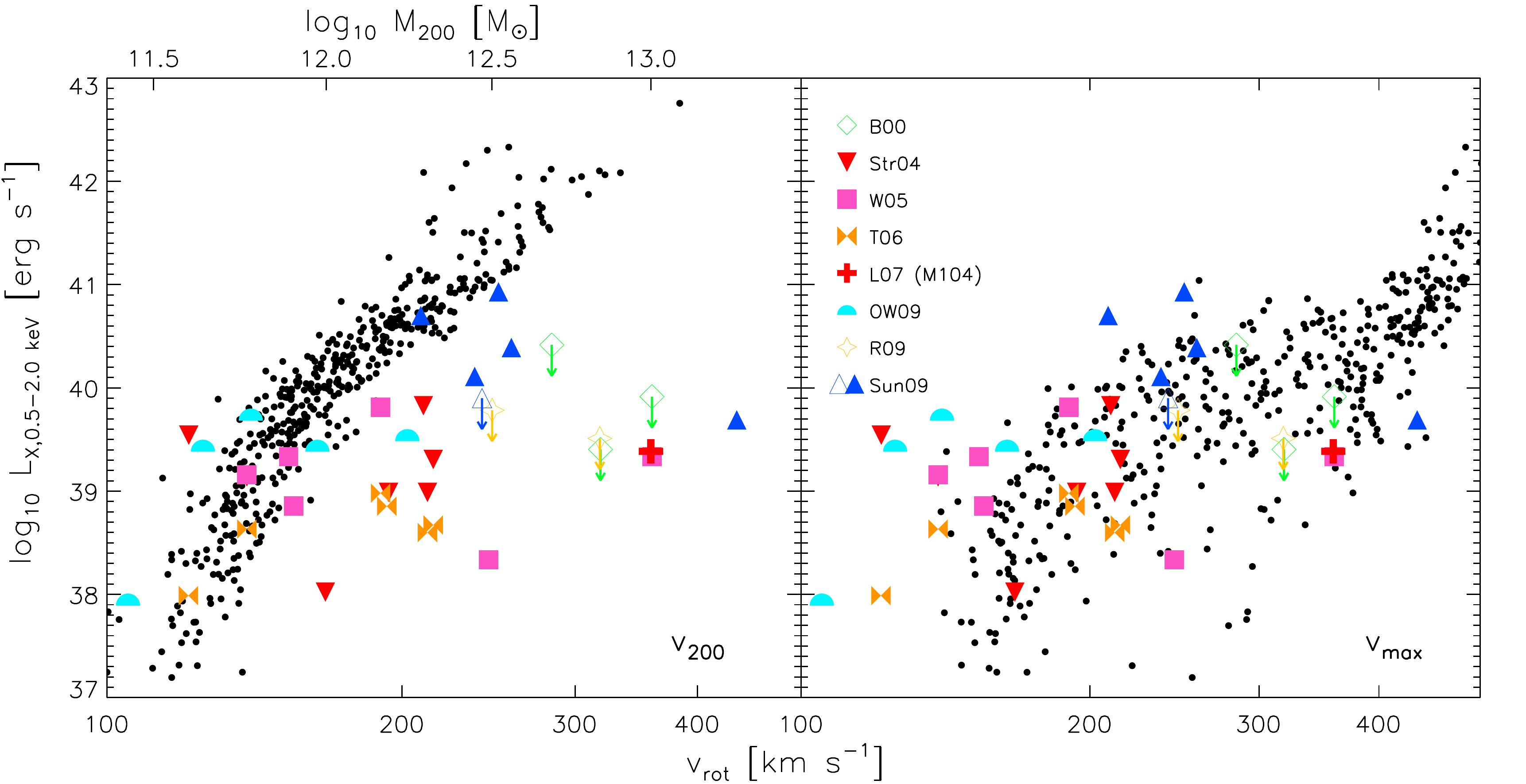}
\caption{The present-day 0.5-2.0 keV X-ray luminosity as a function of disc rotation velocity. In the left-hand panel, we identify the disc rotation speed with $v_{200}$ (the halo circular velocity at $r_{200}$), following WF91's assumption of an isothermal halo density profile. In the right-hand panel, we use instead $v_{\rm max}$, the maximum of the halo radial circular velocity profile. The observational data (various coloured symbols) are the same in both panels; the disc rotation velocities were extracted from the \textsc{HyperLeda} database. Black dots represent galaxies drawn from our simulations. We give the associated mass, $M_{200}$, for a given $v_{200}$ on the upper axis of the left-hand panel, since the conversion between the two is trivial. The data show that the assumption $v_{\rm rot} = v_{200}$ is inaccurate; the approximation $v_{\rm rot} = v_{\rm max}$ produces much better agreement between the simulations and the observational data.}
\label{fig:Lx_v_2up}
\end{figure*}

In this section we examine the X-ray scaling relations found in our hydrodynamic simulations.  In Fig.~\ref{fig:Lx_Lk} we plot the soft (0.5-2.0\keV) X-ray luminosity of the simulated galaxies (black dots), as a function of their $K$-band luminosity.  The colour symbols are measurements for real disc galaxies collated from the literature \citep{Benson_et_al_00,Strickland_et_al_04,Wang_05,Tullmann_et_al_06,Li_Wang_and_Hameed_07,Owen_and_Warwick_09,Rasmussen_et_al_09,Sun_et_al_09}.  Systems for which X-ray detections were made are denoted by solid symbols, whilst those with only upper limits are denoted by open symbols with down arrows. In the case of X-ray detections, we use only the reported extra-planar X-ray luminosity, which has been corrected for point source contamination (by spatial filtering of bright sources and the inclusion of a power-law component in the spectral fitting for faint sources). Where X-ray luminosities have been quoted in other passbands, we correct into the $0.5-2.0\keV$ band using the \textsc{Pimms} software\footnote{http://heasarc.nasa.gov/docs/software/tools/pimms.html} \citep{Mukai_93}. $K$-band luminosities have been extracted from \ipac's online database for the Two Micron All-Sky Survey \citep[\twomass,][]{Skrutskie_et_al_06_short}.

The $K$-band luminosity is a good tracer of stellar mass, which itself is expected to correlate with halo mass. The simulations yield an X-ray vs.\ $K$-band luminosity relation that is in very good agreement with the observed detections, and is consistent with the upper limits for non-detections. It is interesting that the simulations also reproduce the large scatter in the observations; we discuss the origin of this scatter in \S~\ref{sec:Lx_Lk_scatter}.  It should be noted that we have not tuned any element of the simulation subgrid physics, for example, the supernovae wind velocity or mass-loading, in order to reproduce the observed relation. As discussed in C09 \citep[see also][]{Schaye_et_al_10_short}, the adopted feedback parameters were chosen only so as to produce a reasonable match to the peak of the overall star formation history of the Universe at $z \approx 2-3$.

The canonical picture posits that the soft X-ray luminosity of hot coronae increases steeply with the overall mass of the system (i.e. the galaxy and its host dark matter halo). It should therefore be easiest to detect soft X-ray emission from the most massive systems. The virial mass of a halo is commonly approximated by the mass, $M_{200}$, of the sphere of radius $r_{200}$ about the centre of the potential that encloses a mean density of 200 times the critical density for closure. In practice, estimating the total mass is difficult and requires recourse to observational proxies. For disc galaxies, the rotation speed, $v_{\rm rot}$, is often used since, under the assumption that the halo has an isothermal density profile, $\rho \propto r^{-2}$, the circular velocity profile, $v_{\rm c}(r) = [GM(<r)/r]^{1/2}$, is simply a constant at all radii. This enables the rotation speed of a centrifugally supported disc to be equated to the virial circular velocity of the halo, $v_{200}\equiv v_{\rm c}(r_{200})$. However, if the structure of the galaxy-halo system deviates from the isothermal profile, the circular velocity profile is not constant, and the relationship between the rotation velocity of a galaxy and the virial mass of the galaxy's host halo becomes more complicated.

The Navarro-Frenk-White \citep[hereafter NFW,][]{Navarro_Frenk_and_White_95,Navarro_Frenk_and_White_96,Navarro_Frenk_and_White_97} profile, shown to be a near universal density law for dark matter haloes, deviates from the isothermal form at both small and large radii (where asymptotically $\rho \propto r^{-1}$ and $\rho \propto r^{-3}$ respectively). Thus, the circular velocity curve $v_{\rm c}(r)$ of the NFW profile varies with radius. For example, an NFW halo with concentration $c = 10$ exhibits a peak circular velocity, $v_{\rm max} = {\rm max}[v_{\rm c}(r)] = 1.2v_{200}$ at $r \sim 0.2r_{200}$. The density profile of galaxy haloes will be modified by the baryons cooling in the halo \citep[e.g.][]{Blumenthal_et_al_86,Jesseit_Naab_and_Burkert_02,Gnedin_et_al_04,Abadi_et_al_09,Duffy_et_al_10}. In general, cold baryons will be preferentially deposited at the halo centre leading to a steepening of the central gravitational potential (both because of the presence of the baryons and the adiabatic contraction of the dark matter halo), raising the peak of the circular velocity curve to $v_{\rm max}/v_{200} \sim (1.5-2.0)$. These simple theoretical arguments indicate that the halo mass cannot be accurately estimated by equating the rotation speed of galaxy discs with $v_{200}$. Both modelling improvements - switching from isothermal to NFW profiles, and accounting for the effects of baryons - are important, with the latter perhaps being somewhat more so.

In Fig.~\ref{fig:Lx_v_2up} we plot the relationship between the soft X-ray luminosity and two different `measures' of the disc rotation speed for the simulated galaxies. In the left-hand panel, we identify $v_{\rm rot}$ with the circular velocity at the virial radius (as in the isothermal profile); in the right-hand panel we use instead $v_{\rm max}$ computed from all mass components, i.e. gas, stars and dark matter. The observational data are the same in both panels; rotation velocities are inclination-corrected 21-cm measurements extracted from the \textsc{HyperLeda} database\footnote{http://leda.univ-lyon1.fr/} \citep{Paturel_et_al_03}.

When we adopt the inaccurate approximation, $v_{\rm rot} \simeq v_{200}$, the simulated galaxies are much too X-ray luminous at a given value of $v_{\rm rot}$ compared to the observational data.  A similar conclusion was drawn by B00 and others when comparing the analytic predictions of WF91 to observational data.  However, when we use $v_{\rm max}$ as a measure of $v_{\rm rot}$ the simulations are in much closer agreement with the data. This is a much more realistic assumption since $r_{\rm max}$ is typically $< 10\kpc$, comparable to the sizes of the simulated discs\footnote{In principle, the rotation speed of the discs could be calculated directly in the simulations. In practice, this is inaccurate because of small particle number and resolution effects. We defer a discussion of the structure of the discs to a more detailed paper.}.  The right-hand panel of Fig.~\ref{fig:Lx_v_2up} immediately indicates that a significant improvement can be made to analytic estimates of $L_{\rm X}$ simply by adopting a realistic density profile for the halo.  However, this effect alone is insufficient to reconcile the analytic predictions of WF91 with the observations and, as we discuss below in \S4, a reduction in the density of the hot gas halo is also required.

\begin{figure}
\includegraphics[width=\columnwidth]{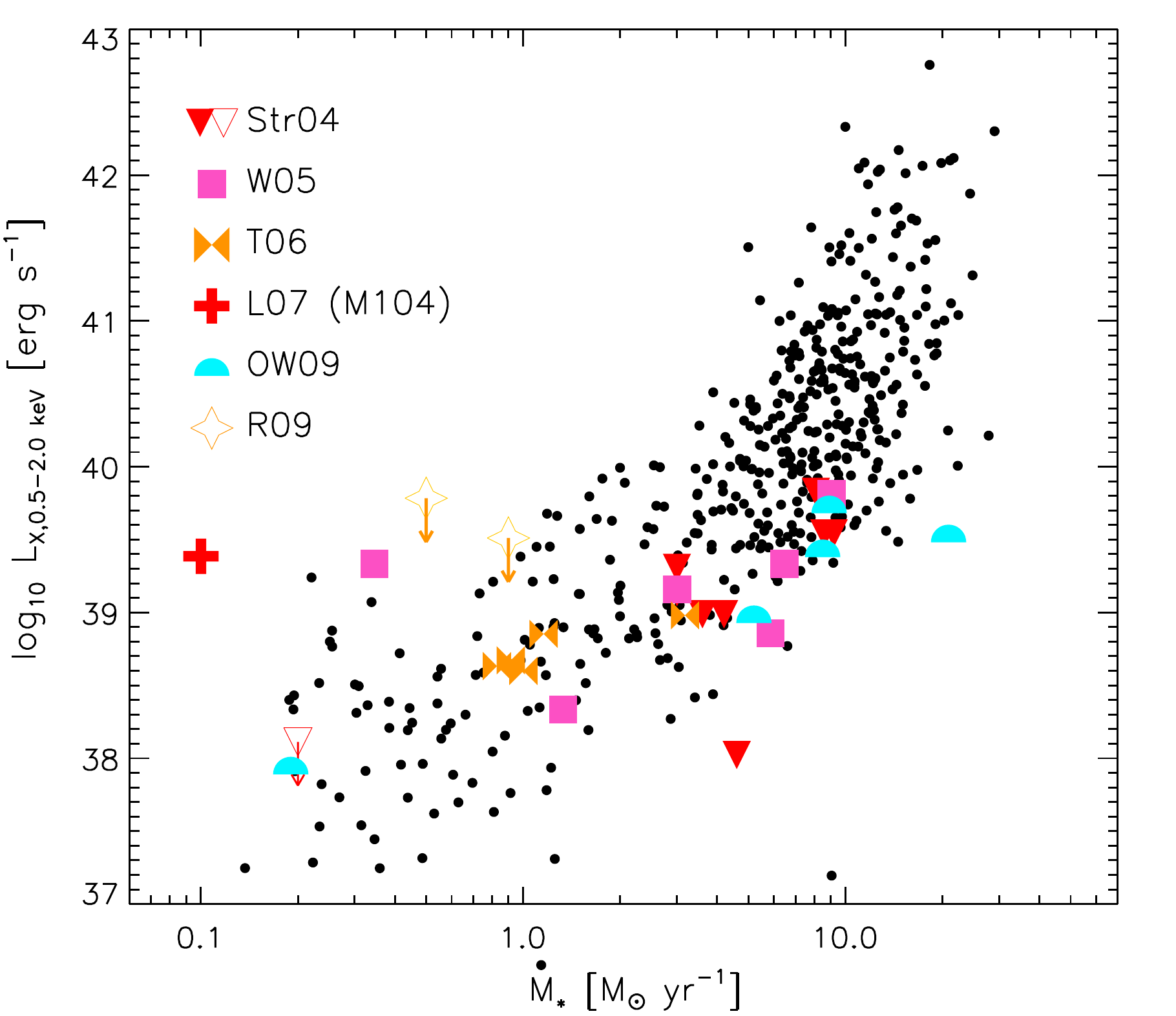}
\caption{The present-day 0.5-2.0 keV X-ray luminosity as a function of star formation rate. The star formation rates of observed galaxies have been derived from near-infrared luminosities, using the relation of \citet{Kennicutt_98}, corrected to our assumed \citet{Chabrier_03} IMF. Note that this is a small correction (see text). The simulated galaxies, shown as black dots, are consistent with both the scaling and scatter of the data.}
\label{fig:Lx_SFR}  
\end{figure}

In several cases where diffuse X-ray emission has been detected from disc galaxies, a correlation between the X-ray luminosity of the corona and the star formation rate of the galaxy has been noted. Furthermore, there exist a few spectacular cases, such as the starburst galaxy M82 \citep{Strickland_et_al_04,Strickland_and_Heckman_07}, where the detected X-ray emission has a biconical morphology. On this basis, several authors have concluded that the X-ray emission must originate from hot \textit{outflowing} gas that has recently been heated by supernovae, rather than from the gravitationally-heated inflowing gas of a cooling flow. In Fig.~\ref{fig:Lx_SFR}, we show the relation between the soft X-ray luminosity and the star formation rate (SFR), $\dot{M}_\star$, of our simulated galaxies. Overplotted are estimates of $\dot{M}_\star$ for several observed systems, obtained from far infrared (FIR) continuum luminosities using the relation of \citet{Kennicutt_98}. For consistency, we correct these SFRs to the \citet{Chabrier_03} IMF assumed in the simulations rather than that adopted by Kennicutt. For this, we assume that our galaxies experience ongoing star formation, and hence that the FIR continuum luminosity is dominated by stars more massive than $5\Msun$. We then scale the inferred SFR by the mass fraction of each IMF comprised by these massive stars (0.23 \& 0.29 for Kennicutt and Chabrier, respectively). For the same FIR continuum luminosity, the inferred SFR corresponding to the Chabrier IMF is therefore 79 per cent of that for the Kennicutt IMF, a relatively small correction.

Whilst a correlation between X-ray luminosity and star formation rate has been found in previous studies that focussed exclusively on galaxies with high X-ray luminosities \citep[e.g.][]{Strickland_et_al_04}, it is interesting to note that the addition of measurements for low X-ray luminosity systems in Fig.~\ref{fig:Lx_SFR} renders the connection much less compelling, even if one excludes outlying systems such as M104. The simulations are broadly compatible with these observations (for $\dot{M}_\star \lesssim 5\Msunyr$), exhibiting a loose correlation between $L_{\rm X}$ and $\dot{M}_\star$ with a scatter in $L_{\rm X}$ of more than a factor of 10 at fixed $\dot{M}_\star$. Interestingly, the simulations indicate an additional reason for caution when interpreting the perceived correlation between $L_{\rm X}$ and $\dot{M}_\star$, in that both quantities are expected to scale with the overall mass of the system. Any correlation between these quantities does not, by itself, demonstrate that the X-ray emission originates in outflows driven by star formation, and we challenge the generality of this frequently made interpretation in \S~\ref{sec:outflow_vs_inflow}.


\subsection{Scatter in the $L_{\rm X}-L_{\rm K}$ relation}
\label{sec:Lx_Lk_scatter}

\begin{figure*}
\includegraphics[width=\textwidth]{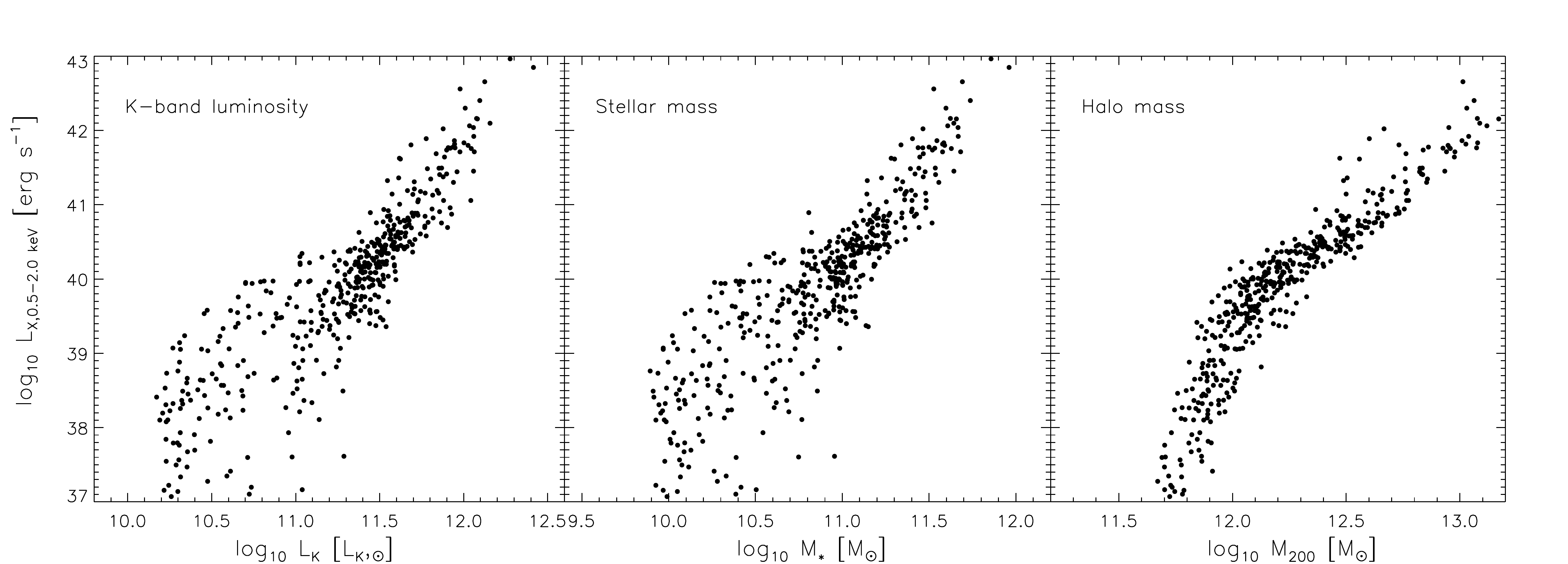}
\caption{The scaling of soft X-ray luminosity with $K$-band luminosity (\textit{left}), stellar mass (\textit{centre}) and halo mass (\textit{right}). The similarity of the left-hand and centre panels demonstrates that the $K$-band luminosity \citep[computed using the \textsc{Galaxev} model of][]{BC03} traces the stellar mass very closely, with a nearly 1:1 relation between $L_{\rm K}$ and $M_\star$.  The contrast in the scatter exhibited in the centre and right-hand panels indicates a significant scatter in stellar mass at fixed virial mass (consistent with the top right panel of Fig.~\ref{fig:baryons_M200}), and it is this effect that drives the scatter in the $L_{\rm X}-L_{\rm K}$ relation.}
\label{fig:scatter_lx_lk}
\end{figure*}

Finally, we consider the system-to-system scatter. The observational scalings show considerable scatter, for example in the $L_{\rm X}-L_{\rm K}$, $L_{\rm X}-v_{\rm rot}$ and $L_{\rm X}-\dot{M}_\star$ relations, which is also present in the simulations. To investigate the origin of this scatter, we plot the scaling between X-ray luminosity and $K$-band luminosity, stellar mass and virial mass, in Fig.~\ref{fig:scatter_lx_lk}. As anticipated, the $K$-band luminosity traces the stellar mass of the galaxy very closely. These panels demonstrate that the scatter in the X-ray scaling relations stems primarily from the scatter in stellar mass at fixed halo virial mass, which itself is sensitive to the assembly history of the galaxy. By contrast, the scatter in $L_{\rm X}$ at fixed halo mass is relatively small for all halo masses considered here, raising the interesting possibility that X-ray properties might be understood in greater detail by focusing on systems for which a more accurate and precise measurement of the halo mass can be found, such as those derived from strong lensing \citep{Ferreras_Saha_and_Williams_05}, planetary nebulae \citep{Romanowsky_et_al_03,Napolitano_et_al_09_short} or the properties of globular clusters \citep{Spitler_and_Forbes_09}.

The scatter in the $L_{\rm X}-L_{\rm K}$ plane for \textit{elliptical} galaxies has received considerable attention, since much larger samples of ellipticals can be compiled. The standard interpretation of the scatter in $L_{\rm X}$ at fixed $L_{\rm K}$ in these samples is, as seen for our sample of disc galaxies, that such galaxies are hosted by dark matter haloes drawn from a wide range of total masses (i.e. $M_{200}$), leading to a similarly wide range of X-ray luminosities \citep{Mathews_and_Brighenti_03,Mathews_et_al_06}. More recently, an analysis of archival \chandra\ data by \citet{Jeltema_Binder_and_Mulchaey_08} uncovered a significantly greater fraction of X-ray detected $L_{\rm K}>L^\star$ galaxies in groups relative to clusters. They concluded that environmental effects, for example ram pressure stripping, are a considerable source of scatter at the bright end of the $L_{\rm X}-L_{\rm K}$ relation. Since we have focussed here on isolated $L^\star$ disc galaxies, the coronae in our sample will not have been affected by ram pressure stripping. Nonetheless, we find considerable scatter in the $L_{\rm X}-L_{\rm K}$ plane, indicating that some fraction of this scatter is due to the range of $M_\star$ at fixed $M_{200}$ (and vice-versa). This is illustrated succintly by the contrasting scatter in the $L_{\rm X}-M_\star$ and $L_{\rm X}-M_{200}$ relations (Fig.~\ref{fig:scatter_lx_lk}, centre and right-hand panels respectively). We note, however, that this does not preclude an effect of the kind found by \citet{Jeltema_Binder_and_Mulchaey_08} for the early-type galaxies hosted by massive clusters in these simulations.

\section{Theoretical interpretation}
\label{sec:theory}

Having established that the \gimic\ simulation generally reproduces the basic scaling relations observed for the soft X-ray luminosity of disc galaxies, it seems unlikely that a fundamental shortcoming in the canonical theory of disc galaxy formation is the reason for its large overprediction of the X-ray luminosities of galaxies. Instead, it seems more likely that inaccuracies in one or more of the approximations adopted by WF91 are to blame. We explore these approximations in this section.

\subsection{The analytic basis for X-ray coronae}
\label{sec:analytics}

In sufficiently massive haloes it is expected that accreting gas
will be shock-heated to a temperature that reflects the depth of the dark
matter potential well, i.e. the halo virial temperature, 
\begin{equation}
  T_{200} = {1\over 3}\,{\mu m_{\rm p}\over k_{\rm B}}\,v_{200}^2 =
  3.6\times 10^5\,{\rm K}\,\left({\mu\over 0.59}\right)\,\left({v_{200}\over 100~{\rm km}\,{\rm
        s}^{-1}}\right)^2.
\end{equation}
In massive haloes, the post-shock cooling time of the gas will, in general, be greater than the dynamical time of the halo, leading to the formation of a quasi-hydrostatic corona. Cooling of some fraction of this hot gas leads to a cooling flow from which a rotationally supported gas disc forms. A stellar disc then grows from quiescent star formation within the cold gas disc. This scenario was first put forward by \citet{White_and_Rees_78} and was implemented in the CDM cosmology by WF91, leading to the important conclusion that massive galaxies should, at the present day, be embedded within extended atmosheres of hot, X-ray luminous gas. Disc galaxies represent the cleanest observational test of this picture because, according to the WF91 model, they are systems with a relatively quiescent star formation history for which the central assumptions of the model appear most justified. These assumptions are as follows:
\begin{itemize}
\item{WF91 assumed that both the dark matter halo and the gas corona follow an isothermal density profile, $\rho \propto r^{-2}$.  The gas profile is normalised to obtain a desired overall gas mass fraction within the virial radius. In the simplest case, where cooling and star formation are inefficient and feedback is energetically incapable of ejecting gas from the halo, $f_{\rm gas}(r_{200}) \approx \Omega_{\rm b}/\Omega_{\rm m}$.  Alternative analytic forms (such as the NFW profile) have also been explored in the context of the WF91 formalism (see, e.g., B00).}
\item{The temperature of the gas is equal to the halo virial temperature, $T_{200}$.}
\item{In the simplest case, the gas is assumed to have a primordial composition of hydrogen and helium ($Y=0.24$). The inclusion of radiative emission due to metal lines increases the predicted X-ray luminosity.}
\item{The gas cools in a single phase quasi-hydrostatic cooling flow at a rate, $\dot{M}_{\rm cool}$, that is determined by the assumed gas density, temperature, and metallicity distributions (see Eqns.~20-22 of WF91). This leads to an X-ray luminosity, $L_{\rm X} \propto \dot{M}_{\rm cool}v_{\rm 200}^2$ \citep[as first highlighted by][]{Thomas_et_al_86}.}
\end{itemize}

These assumptions enable the complex physical processes underlying galaxy formation to be condensed into a relatively simple set of analytic equations. We expect that, to some degree, all of these assumptions will be violated in the real world and in cosmological hydrodynamic simulations. In the following subsection, we quantify to extent to which the analytical approximations are violated in the simulations and the impact on the predicted X-ray emission from massive disc galaxies.

\subsection{Why are the predicted X-ray luminosities now in agreement with observations?}

\begin{figure*}
\includegraphics[width=0.9\textwidth]{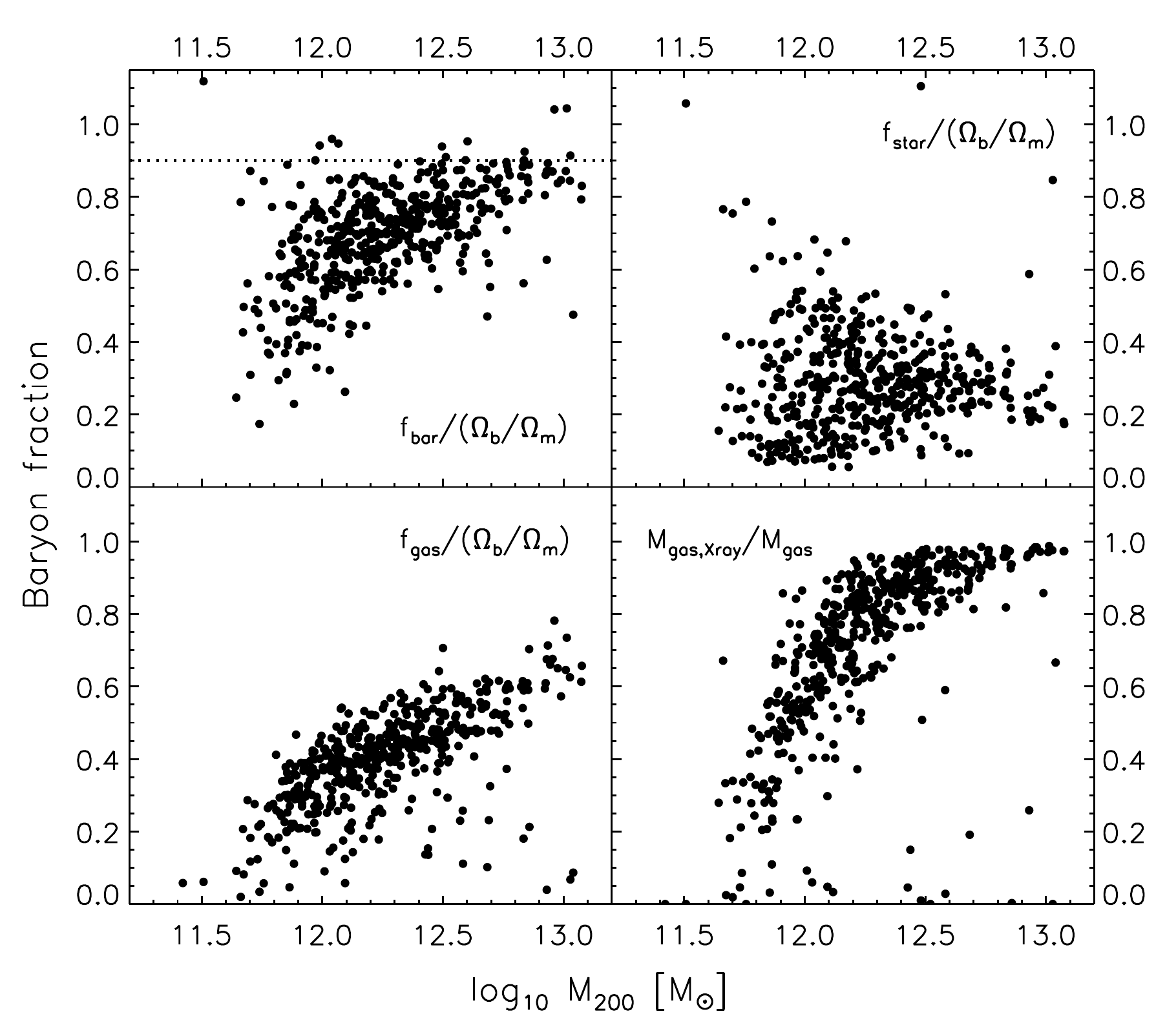}
\caption{The baryon content, at $z=0$, of the haloes included in the simulated galaxy sample. The total baryon fraction, $f_{\rm b}\equiv (M_\star+M_{\rm gas})/M_{200}$ (\textit{top left}), the stellar fraction $f_\star\equiv M_\star/M_{200}$ (\textit{top right}), and the gas fraction $f_{\rm gas}\equiv M_{\rm gas}/M_{200}$ (\textit{bottom left}) are normalised by the cosmic baryon fraction of $\Omega_{\rm b}/\Omega_{\rm m}$. The bottom right panel shows the mass fraction of gas that is hot ($T>2\times10^5$ K), and thus potentially X-ray luminous. The dotted line in the top-left panel marks the baryon fraction found by \citet{Frenk_et_al_96} and \citet{Crain_et_al_07} in the case of non-radiative gas.} 
\label{fig:baryons_M200}
\end{figure*} 

\citet{Toft_et_al_02} concluded that the X-ray luminosity of galaxies in their hydrodynamic simulations was lower than in analytic models because efficient radiative cooling over cosmic history substantially reduced the hot gas fraction of dark matter haloes by $z=0$. On the other hand, C09 demonstrated that supernova winds strongly reduce the mass of dense, highly X-ray luminous gas near the centres of haloes with circular velocity $v_{200}\lesssim 300\kms$ at $z=0$. Thus, both cooling and heating processes can act to reduce the X-ray luminosity of galaxies.

We begin by scrutinising the halo baryon content of our galaxy sample in greater detail, in Fig.~\ref{fig:baryons_M200}. The first three panels show the overall baryon fraction (gas + stars, \textit{top left}), the stellar fraction (\textit{top right}) and the gas fraction (\textit{bottom left}), each normalised by the cosmic baryon fraction. The final panel (\textit{bottom right}) shows the fraction of halo gas that is X-ray luminous (i.e. $T \ge 2\times 10^5\K$). As a consequence of supernova feedback, the overall baryon fraction is a strong function of halo mass: haloes with $M_{200} \lesssim 10^{12}\Msun$ have overall baryon fractions that are typically half of the cosmic value. At higher masses, halo baryon fractions approach the $f_{\rm b} \simeq 0.9\Omega_{\rm b}/\Omega_{\rm m}$ limit found by \citet{Frenk_et_al_96} and \citet{Crain_et_al_07} in the non-radiative regime, indicated in the top left panel by a dotted line. The most massive haloes in our sample, whose masses approach those of small galaxy groups, are therefore essentially `baryonically closed' systems\footnote{We note here that this result violates observational constraints on the galaxy group scale derived from a combination of optical and X-ray measures, but this can be understood in terms of the much greater role expected to be played by feedback from active galactic nuclei (AGN) in this high-mass regime \citep[e.g.][]{McCarthy_et_al_09_short}. We do not attempt to model this particular feedback mechanism in these simulations.}. However, for all systems (including the most massive ones) a non-negligible fraction of the baryons that reside within the virial radius are in the form of stars (top-right panel of the figure) and, in the case of low mass haloes, cold gas (lower panels). The stellar mass fraction in haloes with mass $M_{200}\lesssim 10^{12}\Msun$ is, on average, similar to the fraction found by \citet{Guo_et_al_09} for galaxies in the Sloan survey; for larger haloes, the stellar masses are about a factor of two too large, but on these scales our neglect of AGN feedback could be significant.

The low mass haloes in our simulated sample have baryon mass fractions that have been reduced relative to the cosmic mean by feedback. The mean gas density in these haloes is also lower than assumed in the canonical picture. This deficit of hot gas immediately points to a likely cause for the lower X-ray luminosities found here compared to WF91. However, even those systems that we identified as baryonically closed, and that therefore still retain relatively high mean densities, exhibit significantly lower X-ray luminosities than predicted analytically. Such low X-ray luminosities warrant an explanation.

\begin{figure*}
\includegraphics[width=\columnwidth]{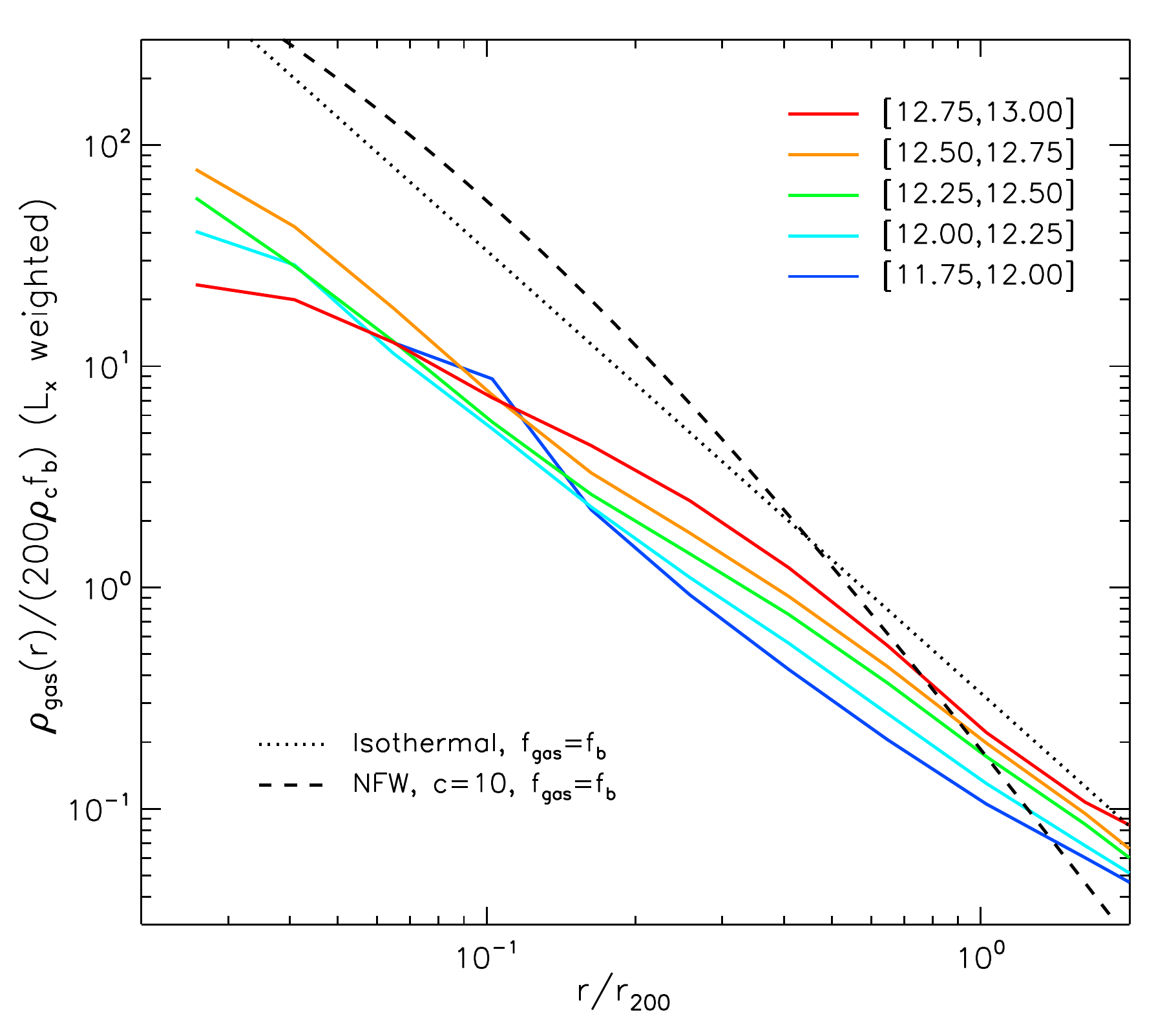}
\includegraphics[width=\columnwidth]{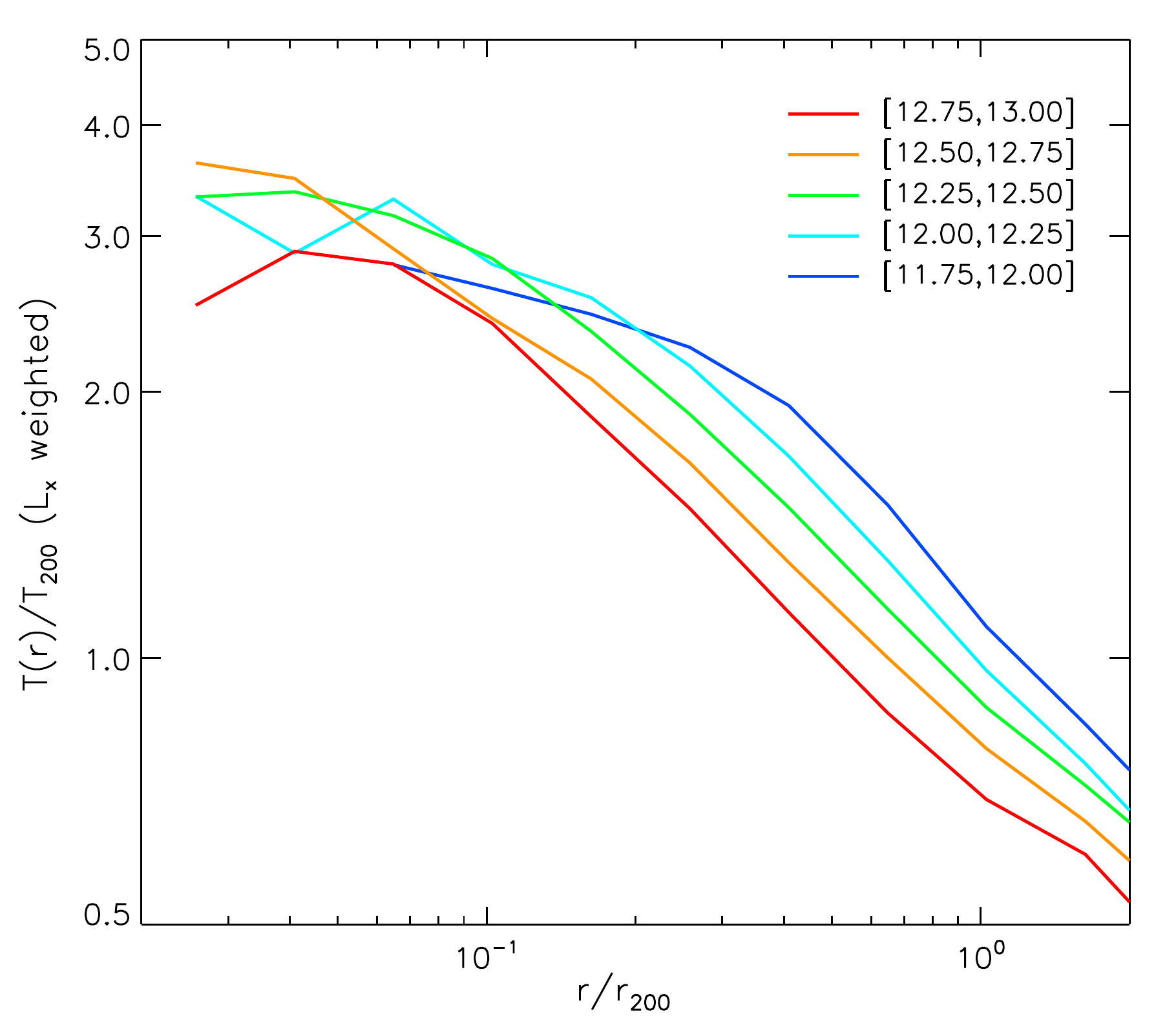}
\includegraphics[width=\columnwidth]{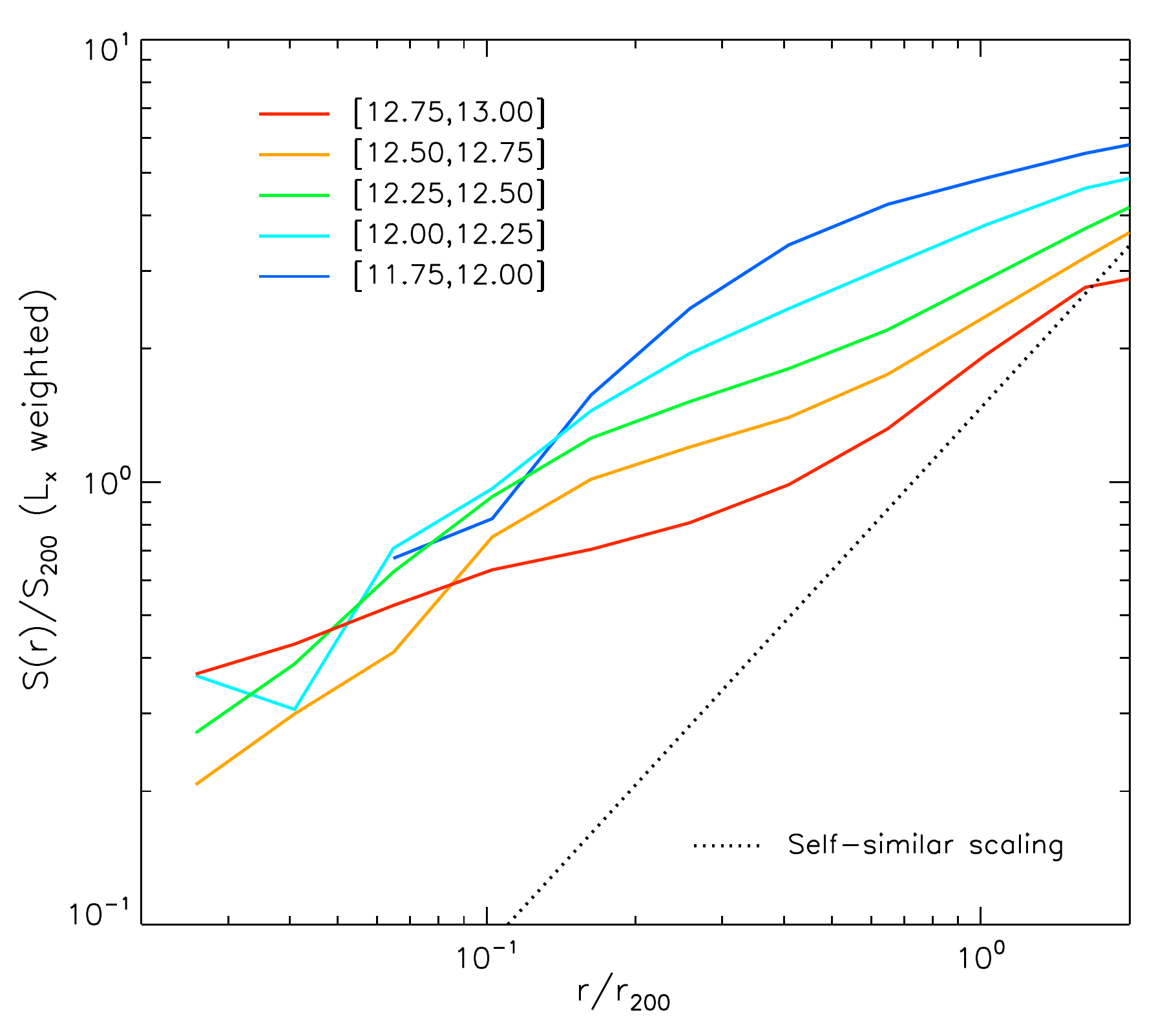}
\includegraphics[width=\columnwidth]{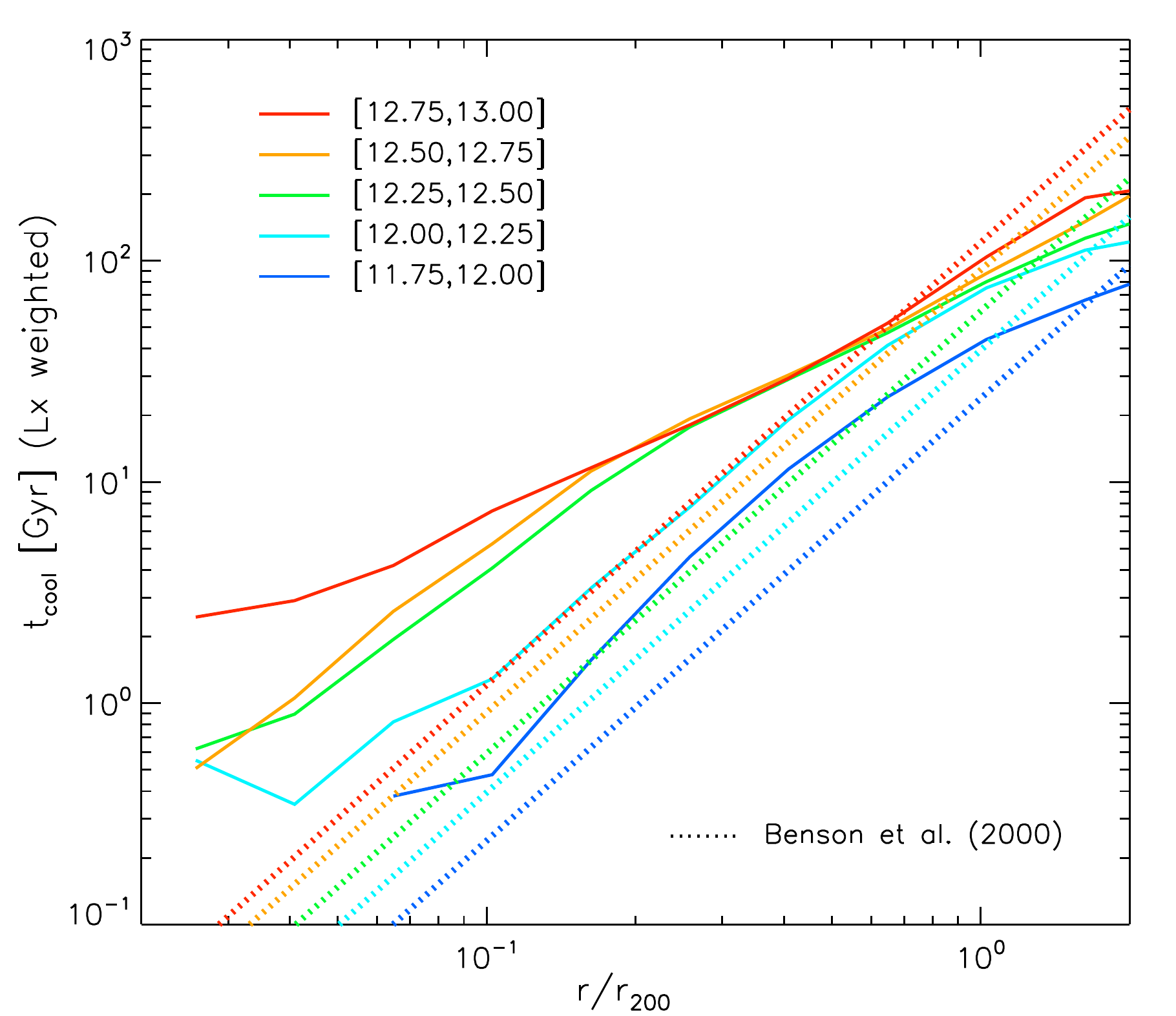}
\caption{Present-day luminosity weighted, three-dimensional, spherically averaged radial profiles of the hot gas bound to the haloes of disc galaxies in our sample. Galaxies are binned in $\log_{10}M_{200}$ at $z=0$ (see legend, units are $\Msun$). Shown are the density (\textit{top left}), temperature (\textit{top right}), entropy (\textit{bottom left}), and cooling time (\textit{bottom right}). The density profiles clearly deviate from the isothermal (\textit{dotted line}) and NFW (\textit{dashed line}) profiles, commonly adopted in analytic models. Feedback from supernova-driven winds drives gas from the dense halo centre, strongly suppressing the X-ray luminosity, particularly for less massive systems. Feedback and non-gravitational heating also render the temperature profile super-virial. The net effect is that the entropy profiles deviate significantly from the self-similar scaling expected in the non-radiative regime, shown as a dotted line in the bottom-left panel. This entropy elevation lengthens the cooling time relative to that expected from analytic modelling under the assumption of primordial composition gas. This is seen in the bottom-right panel where the dotted lines correspond to the canonical WF91 model as calculated by \citet{Benson_et_al_00}.}
\label{fig:profiles}
\end{figure*}

A clearer understanding can be obtained from radial profiles of the thermodynamic state of the hot gas, rather than just the mass fraction (since X-ray luminosity exhibits a strong dependence on the density distribution and is particularly sensitive to its profile at the halo centre). Spherically-averaged radial profiles of the key thermodynamic quantities at $z=0$ for all gas with $T \ge 2\times 10^5\K$ are shown in Fig.~\ref{fig:profiles}: density (\textit{top left}), temperature (\textit{top right}), entropy (\textit{bottom left}) and cooling time (\textit{bottom right}). The profiles are split into halo mass bins and the plotted lines represent the median profiles of haloes in each bin. The profiles are discussed in turn below.

\begin{itemize}
\item \textit{Density.} The gas density profiles of the simulated haloes deviate significantly from the favoured analytic forms, isothermal (\textit{dotted line}) and NFW (\textit{dashed line}). The least massive galaxies exhibit the greatest deviation from the analytic models. Interestingly, at large radii where the cooling time of the gas is relatively long (see bottom right panel), the profiles show a monotonic trend in mass, reflecting an increasing deviation from the analytic profiles in less massive haloes. Suppression of the hot gas density, by cooling (or more specifically, star formation, since this is the terminal state for cooled baryons) and feedback, is therefore most pronounced in these systems. Most importantly, the profiles indicate that star formation and feedback preferentially suppress the hot gas density at the \textit{centre} of haloes, and it is the central density to which the X-ray luminosity is most sensitive. We note that, typically, half of the total X-ray luminosity of a simulated galaxy comes from gas within the central 10 per cent of the virial radius. 

\item \textit{Temperature.} The gas at the halo centre is not isothermal, but shows a mild negative temperature gradient. Gas temperatures reach up to 3 times the virial value in the central regions and fall slightly below it at the virial radius. The origin of central super-virial temperatures is twofold: firstly, heating by feedback and, secondly, gravitational compression that results from the steepening of the halo's central potential as it accumulates cold baryons. While this illustrates that the assumption that the gas is at the virial temperature is incorrect in detail, the error incurred by adopting this approximation is not large. Typically, we find that it affects the luminosity by no more than 30 per cent, and it exhibits a mild dependence on the heavy element abundances. This stems from the fact that the virial temperature of a halo with $M_{200} \sim 10^{12}\Msun$ is $\sim 10^6\K$, which roughly corresponds to the minimum of the cooling function. Thus, a change in $T$ by a factor of a few yields only a small change in $\Lambda(T,Z)$.  

\item \textit{Entropy.} The density and temperature profiles reflect the underlying entropy configuration and the depth of the gravitational potential well (which is dominated by dark matter). Unlike the density or temperature, which are sensitive to processes that compress or rarify the gas, entropy is conserved in any adiabatic process. Non-adiabatic cooling and heating processes are therefore most clearly described by entropy, since cooling \textit{always} reduces the local entropy of a system, whilst heating \textit{always} raises it. 

Entropy therefore maintains a record of the thermodynamic history of the gas \citep[e.g.][]{Voit_et_al_03,Voit_et_al_05}. We consider here the`virial entropy', which is defined as
\begin{equation}
S_{200} = \frac{k_{\rm B}T_{200}}{[n_{\rm e,200}]^{2/3}}.
\end{equation}
As can be clearly seen in the bottom-left panel of Fig.~\ref{fig:profiles}, the hot gas in our simulated haloes exhibits a very strong deviation from the self-similar entropy scaling\footnote{The dotted line represents a power-law fit to the entropy profiles of a sample of galaxy groups and clusters simulated by \citet{Voit_Kay_and_Bryan_05} with non-radiative hydrodynamics. Over the range $0.1-1.0r_{200}$, hydrodynamic simulations employing either adaptive mesh refinement (AMR) or SPH algorithms yield consistent profiles.} of \citet[][\textit{dotted line}]{Voit_Kay_and_Bryan_05} which is the entropy distribution expected in the case where cooling and feedback are negligible. The `excess entropy' with respect to the self-similar scaling - for which there is already firm observational evidence on the scale of systems with $M_{200} > 10^{13}\Msun$ \citep[e.g.][]{Balogh_Babul_and_Patton_99,Johnson_Ponman_and_Finoguenov_09,Sun_et_al_09} - therefore represents a strong indication that non-gravitational processes are vitally important in establishing the thermodynamic properties of hot galactic coronae. 

\item \textit{Cooling time.} As a result of the increased entropy (and, in turn, reduced gas density), the cooling time of the hot gas in the simulated galaxies is longer than in the WF91 model. Therefore, if cooling dominates non-gravitational heating and there is a net inflow of hot gas (see \S 4.3 below), the relatively long cooling time of the gas compared to the halo dynamical time (typically $\sim 1-2$ Gyr) implies that the flow is quasi-hydrostatic, as assumed by WF91. We note that the cooling times we compute for the simulated galaxies are self-consistent, in that they account for the thermodynamics and chemical abundances of gas particles. (Typically, we find the mean emission-weighted metallicity of the hot gas to range between $0.3-1.3\Zsun$.) The profiles are compared with the analytic counterparts proposed by WF91 for isothermal gas of primordial composition (dotted lines). The discrepancy between the simulations and the WF91 model would be greater if metal line emission was taken into account in the latter.
\end{itemize}

\begin{figure} 
\includegraphics[width=\columnwidth]{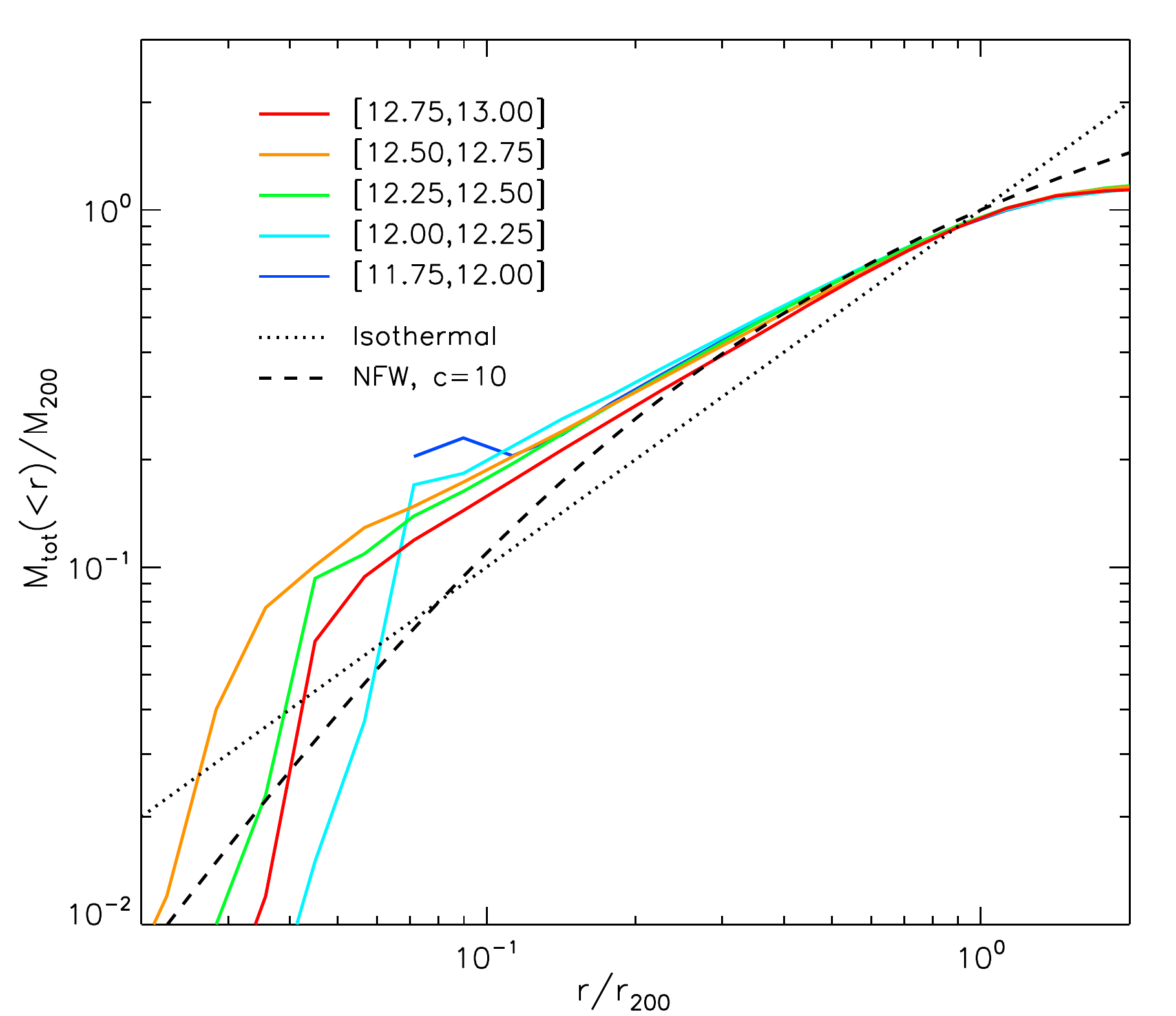} 
\caption{Present-day three-dimensional, spherically-averaged cumulative radial profile of \textit{all} mass (i.e. gas, stars and dark matter) comprising the main subhalo of disc galaxies included in our sample. As per Fig.~\ref{fig:profiles}, galaxies are binned in $\log_{10}M_{200}$ at $z=0$. The profiles exhibit excellent agreement with the NFW (\textit{dashed line}) profile at large radii, but are more concentrated than both the isothermal (\textit{dotted line}) and NFW profiles within $\sim0.3r_{200}$ due to the accumulation of cooled baryons.}
\label{fig:total_mass_profile}
\end{figure}

For reference, we show in Fig.~\ref{fig:total_mass_profile} the cumulative total mass radial profiles (i.e. including gas, stars and dark matter) of the selected systems, again as a function of halo mass. The profiles trace the NFW form (\textit{dashed line}) closely at large radii, but for $r\lesssim0.3r_{200}$ they are more concentrated than both the NFW form (for a reasonable choice of the concentration parameter) and the isothermal form (\textit{dotted line}), down to the innermost radius reliably probed given the resolution of the simulations, which can be approximated as $\sim4$ smoothing lengths \citep{Hernquist_and_Katz_89}, corresponding to a few percent of $r_{200}$. The boost in concentration for the central mass profile follows from the accumulation of cooled baryons that also gives rise to the mildly super-virial gas temperatures seen in Fig.~\ref{fig:profiles}, and is broadly consistent with the picture of baryon-induced halo profile backreactions recently discussed by \citet{Duffy_et_al_10}.

\begin{figure*}
\includegraphics[width=\columnwidth]{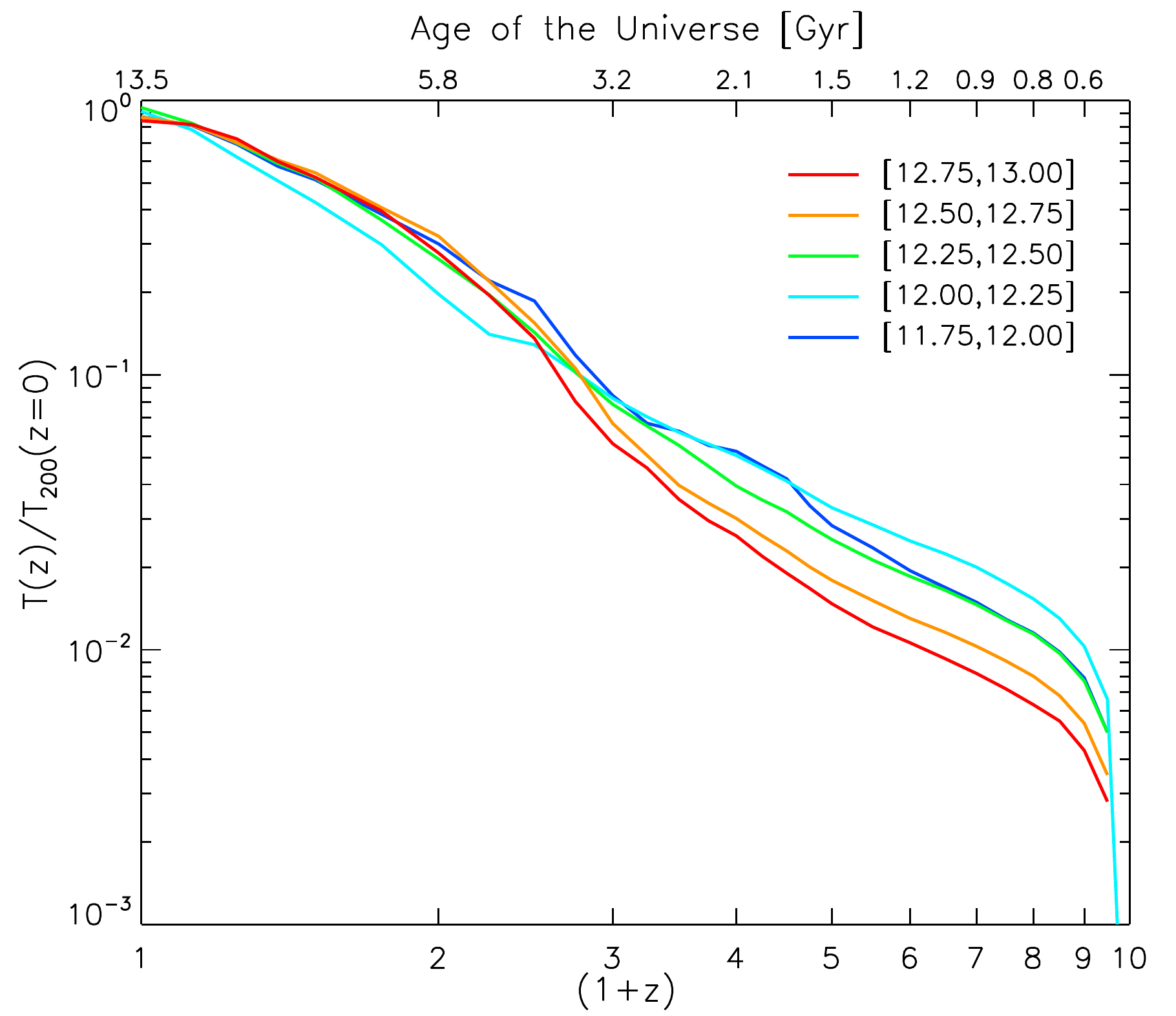}
\includegraphics[width=\columnwidth]{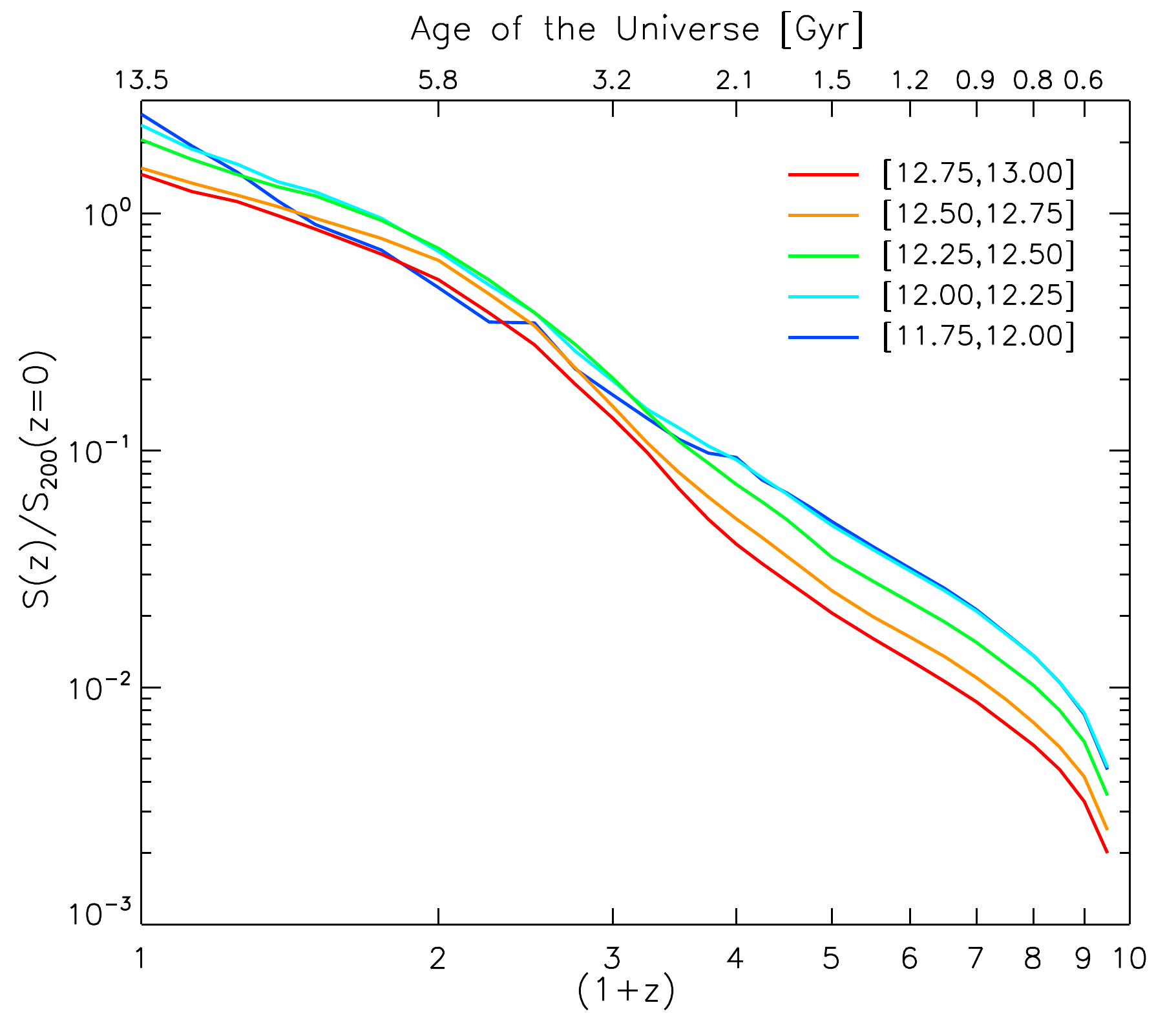}
\caption{The median temperature and entropy histories of gas that resides in the hot coronae of our simulated $\simeq L^\star$ disc galaxies at $z=0$. Galaxies are binned as per Fig.~\ref{fig:profiles}. The profiles are normalised to the present-day virial temperature and entropy, respectively, of each system ($S_{200} \propto T_{200} \propto M_{200}^{2/3}$). The rapid elevation of both quantities at $z=9$ signals the onset of \textsc{Hi} reionisation. Both quantities steadily rise to $z=0$, experiencing the most rapid increase over the interval $z\simeq1-3$, corresponding to the peak in the star formation history of the simulations (see C09).} 
\label{fig:history}
\end{figure*}

To summarise, the X-ray luminosity of galactic coronae is low primarily because the central density of hot gas has been reduced. This, in turn, can be interpreted in terms of a raised adiabat (entropy) for the gas. It is important to establish when the entropy was raised and, if possible, by what mechanism. Fig.~\ref{fig:history} shows the median temperature (\textit{left-hand panel}) and entropy (\textit{right-hand panel}) histories of all gas that is in the X-ray luminous phase at $z=0$, binned by halo mass (also at $z=0$). The sharp rise in both quantities at $z=9$ is caused by the onset of \textsc{Hi} reionisation, which is assumed to occur instantly and uniformly. In absolute terms, however, the entropy and thermal energy injected into the gas from reionisation are small compared to the final entropy and temperature of the hot gas. However, a second sharp rise in both quantities is evident over the epoch $1 \lesssim z \lesssim 3$, corresponding to the peak in $\dot{M}_\star$ for our galaxies (which we investigate in greater detail in \S~\ref{sec:sfh_hot_cold}). Unfortunately, this rise alone does not enable us to distinguish between star formation and feedback as the primary mechanism by which the median entropy of hot gas is raised, since feedback is intimately tied to star formation, both spatially and temporally (both in real galaxies and our simulations).

\begin{figure}
\includegraphics[width=\columnwidth]{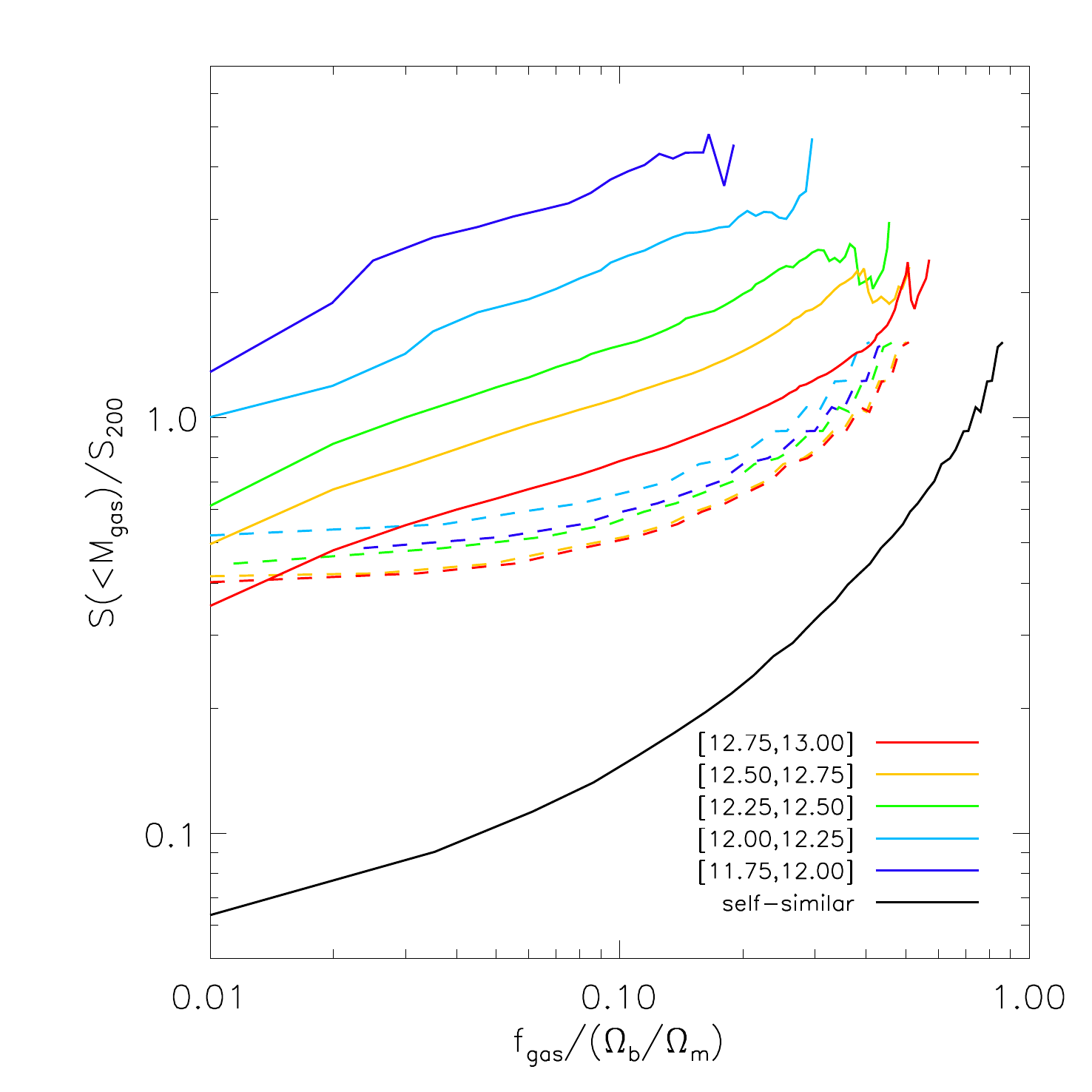}
\caption{The sorted gas entropy as a function of enclosed gas mass fraction (solid curves). This is calculated for each galaxy by ordering the hot gas halo particles by their entropy and plotting the entropy, $S$, against the total mass of gas of particles with entropy lower than $S$.  The solid black curve represents the self-similar result obtained from non-radiative cosmological simulations \citep[see][]{Voit_Kay_and_Bryan_05}.  The dashed curves represent the expected $S(<M_{\rm gas})$ distribution after cooling has removed the lowest entropy gas from the hot phase, which is calculated by removing from the self-similar distribution all gas with entropy that encloses $f_{\rm gas} = f_{\rm star}$ (see text). The similarity of the dashed and solid curves for high halo masses implies that radiative cooling/star formation is primarily responsible for `raising' the entropy of the (remaining) hot gas. For low mass haloes, however, feedback is also required to raise the adiabat.  This mode of entropy production dominates for systems with masses similar to the Milky Way and lower.} 
\label{fig:entropy_mgas}
\end{figure}

In order to obtain insight into which physical mechanism is primarily responsible for lowering the central gas density of the simulated disc galaxies, we plot in Fig.~\ref{fig:entropy_mgas} the sorted gas entropy as a function of enclosed gas mass.  This is calculated for each galaxy by ordering the hot gas halo particles by their entropy and plotting the entropy, $S$, against the total mass of gas of particles with entropy lower than $S$.  In this diagram, cooling selectively removes the lowest entropy gas from the hot phase (converting it into cold gas and stars), which results in a shift in the $S(<M_{\rm gas})$ curves to the left \citep[see][]{Bryan_00,Voit_et_al_02}. Feedback, by contrast, actually heats the gas (shifts the curves upward) but, if it is energetic enough, it may also eject gas from the system (shift to the left).

The solid coloured curves in Fig.~\ref{fig:entropy_mgas} represent the median $S(<M_{\rm gas})$ profiles for various mass bins.  The solid black curve represents the self-similar result obtained from non-radiative cosmological simulations \citep[see][]{Voit_Kay_and_Bryan_05}. To determine whether radiative cooling/star formation is the dominant process, we truncate the self-similar $S(<M_{\rm gas})$ distribution (i.e., shift to the left) using the calculated stellar fractions from the simulated galaxies (dashed curves).  This is done by removing all gas with entropies lower than the entropy that encloses $f_{\rm gas} = f_{\rm star}$.  (The remaining gas mass fraction is then just the original gas mass fraction minus the stellar mass fraction.)  This procedure yields the expected $S(<M_{\rm gas})$ distribution after cooling has removed the lowest entropy gas from the hot phase\footnote{More precisely, this represents an {\it upper limit} to the entropy distribution after cooling has selectively removed the lowest entropy gas.  It represents an upper limit as it has been implicitly assumed that any gas that has not cooled out of the hot phase flows adiabatically inward to replace the gas that has cooled out.  In reality, however, some of the remaining hot gas will have its entropy reduced by cooling.}.

A comparison between the dashed and solid curves demonstrates that it is only for the most massive haloes in our sample that the truncated self-similar distribution approaches the actual $S(<M_{\rm gas})$ distribution in our simulations.  For these systems, therefore, radiative cooling/star formation alone is sufficient to explain the high entropy of the halo gas.  For lower mass systems, on the other hand, selectively removing the lowest entropy gas does not by itself raise the entropy of the hot gas sufficiently to reproduce the simulated $S(<M_{\rm gas})$ distribution.  For a halo with $M_{200} \sim 10^{12}$ M$_\odot$, for example, the central entropy (within, say, $f_{\rm gas}/(\Omega_{\rm b}/\Omega_{\rm m}) = 0.1$) can be raised by a factor of $\approx5$ by radiative cooling/star formation, but another factor $\approx5$ is required to match the simulations.  Thus, for galaxies with masses similar to the Milky Way and lower, feedback (at $z\approx1-3$) begins to dominate entropy production.  This is consistent with the reduced overall baryon mass fractions of haloes with this mass and lower seen in the top-left panel of Fig.~\ref{fig:baryons_M200}.  Note that feedback from accreting supermassive black holes, which is neglected in the present study, would presumably reduce the importance of the cooling mechanism for the more massive systems.

\subsection{The physical nature of X-ray-emitting gas}

We have demonstrated that star formation (via radiative cooling) and, more importantly, feedback from star formation, particularly during the epoch corresponding to the peak in the cosmic star formation rate density ($1\lesssim z \lesssim 3$), determine the state of the high entropy, low density X-ray luminous coronae of Milky Way-like galaxies in the \gimic\ simulations.  This contrasts with the picture posited by WF91, in which the dense corona that forms around a galaxy is assumed to be isothermal and to trace the dark matter distribution. In their model a cooling flow is established which fuels the ongoing star formation necessary to replenish stellar mass lost to the ISM by stellar evolution, and to offset the transfer of mass from discs to bulges that results from mergers and bar instabilities. As seen in Fig.~\ref{fig:Lx_SFR}, despite having hot gas reservoirs that are less dense and have longer cooling times than assumed in the WF91 model, our sample of galaxies still exhibits ongoing star formation. This raises two key questions regarding the coronae of the simulated galaxies: i) is the hot gas of the corona inflowing (accreting) or outflowing (in a wind), and ii) are galaxy discs replenished by ongoing star formation fed by the accretion of hot or cold gas? We explore these questions in order.

\subsubsection{X-ray luminous gas: inflowing or outflowing?}
\label{sec:outflow_vs_inflow}

Since the median temperature and entropy histories shown in Fig.~\ref{fig:history} evolve smoothly, one might na\"ively rule out the hypothesis that the majority of the gas surrounding our simulated galaxies was recently deposited into the hot, X-ray luminous phase by supernova-driven winds. To assess this hypothesis more precisely, however, we resort to a more direct test of the inflow/outflow scenario.

It is possible to measure the fraction of X-ray luminous gas in outflows because the simulation code tracks when, if ever, a baryonic particle last attained a density greater than the threshold for star formation ($n_{\rm H} = 0.1\cmcubed$). Particles with density greater than this threshold are considered part of the ISM, and are subject to an imposed equation of state (EOS), $P\propto\rho^{\gamma_{\rm EOS}}$, where our choice of $\gamma_{\rm EOS} = 4/3$ ensures that both the Jeans length and the ratio of the SPH kernel to the Jeans length are independent of density \citep{Schaye_and_Dalla_Vecchia_08}. In order to transit into the hot phase, ISM particles (i.e., those on the EOS) must have formed part of an outflow driven by winds.

\begin{figure*}
\includegraphics[width=\columnwidth]{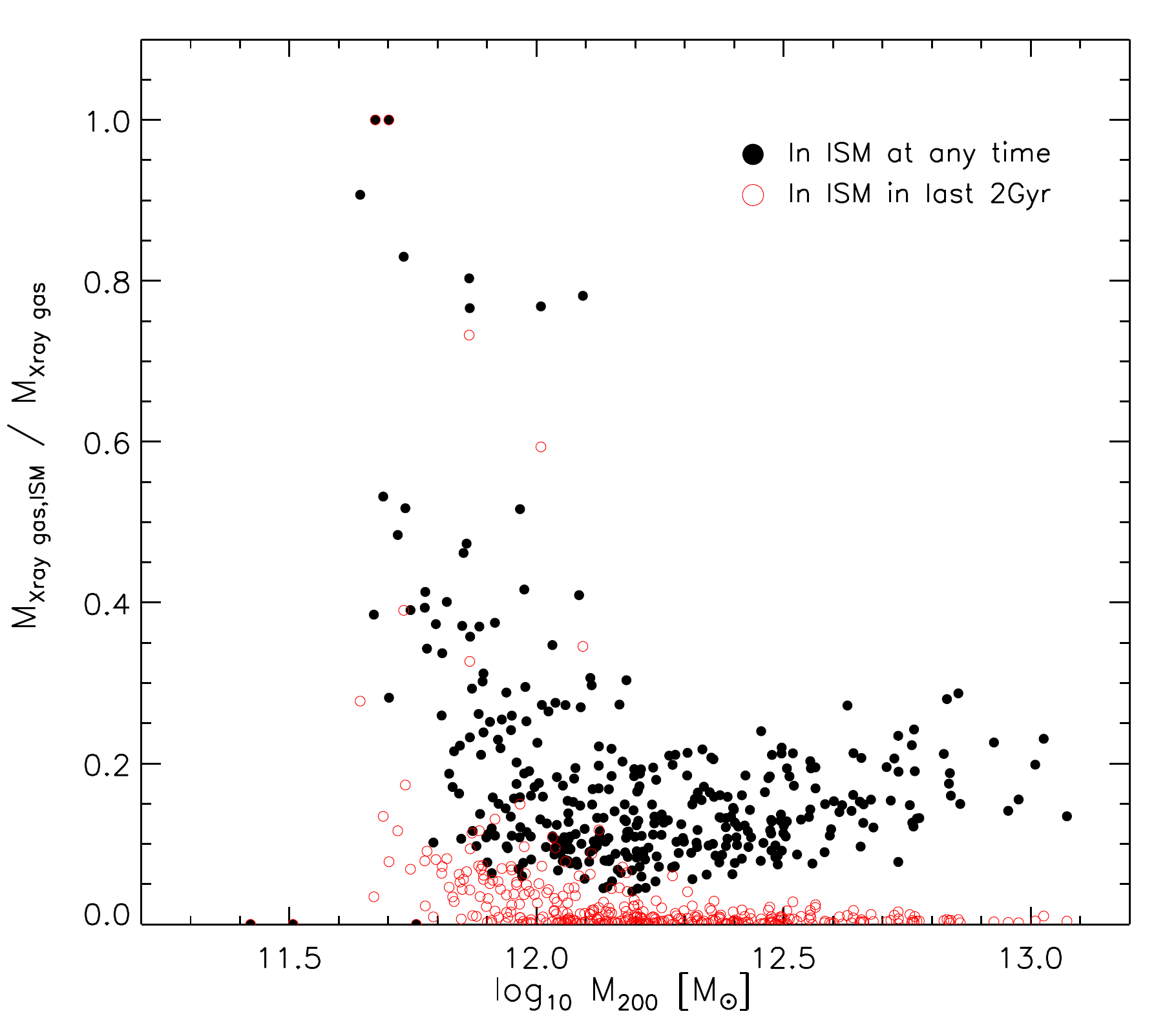}
\includegraphics[width=\columnwidth]{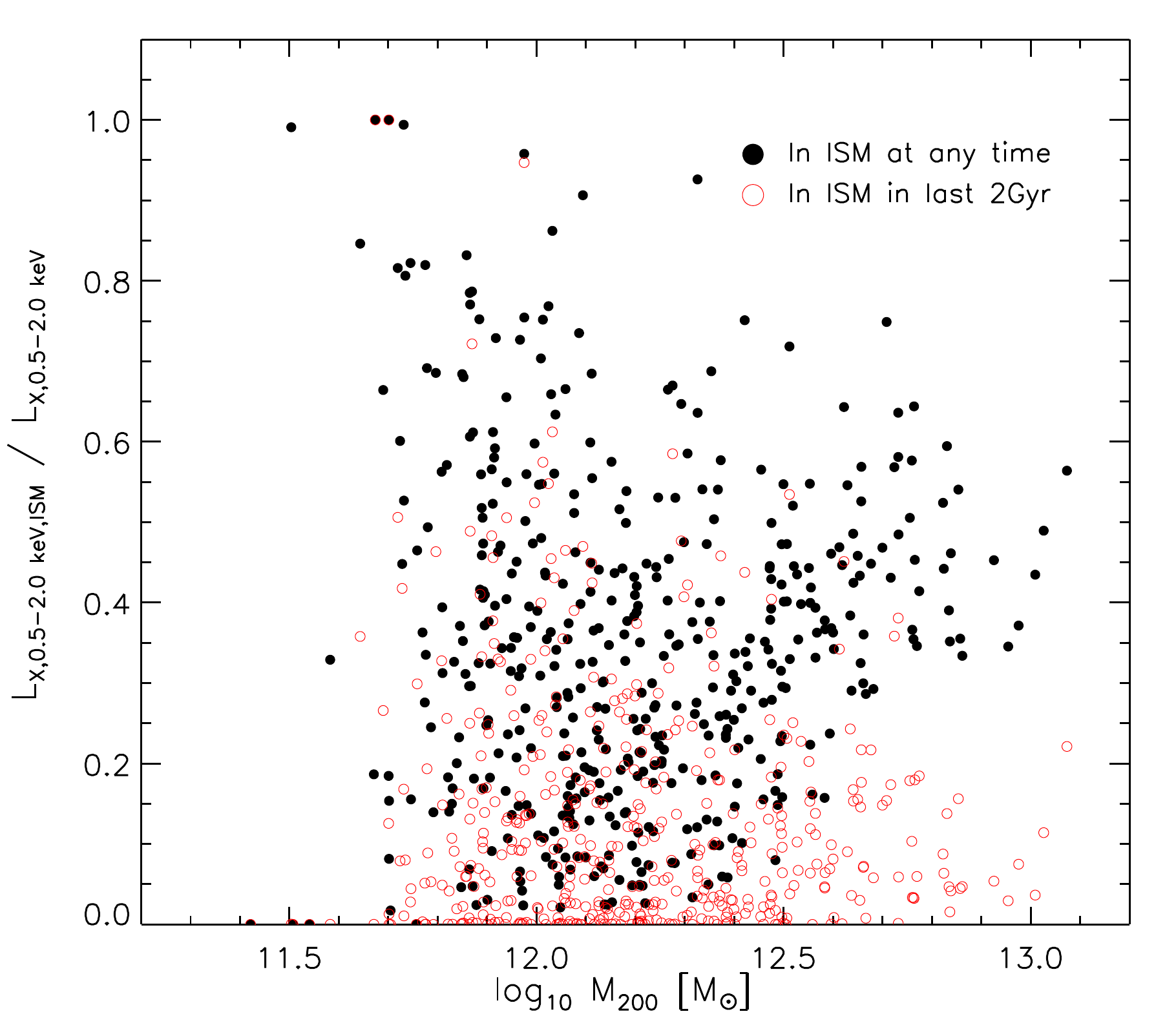}
\caption{The fraction of X-ray luminous mass (\textit{left-hand panel}) and soft X-ray luminosity (\textit{right-hand panel}) contributed by gas in our simulated galaxies that has previously been in the star-forming `equation of state' (EOS) phase that mimics the thermodynamics of the ISM. By definition, transit from the ISM to the hot phase is via an outflow. Each galaxy contributes one black dot and one red circle to each plot; black dots denote the fractions due to gas that has been in the ISM at any time in the past, whilst red circles show the fractions for gas that has been in the ISM in the last 2\Gyr, corresponding to a dynamical time for $L^\star$ galaxies. Above $M_{200} \approx 10^{12}\Msun$, only $10-20\%$ of the hot gas mass has {\it ever} been processed through the ISM, and only a few per cent has been within the last 2\Gyr. However, this gas contributes a disproportionately high fraction of the coronal X-ray luminosity, owing to its centrally concentrated distribution and relatively high metallicity.} 
\label{fig:on_eos}
\end{figure*}

The left-hand panel of Fig.~\ref{fig:on_eos} shows the mass fraction of presently X-ray luminous gas that has been in the ISM, i) at any time in the past (\textit{black dots}), and ii) within the last 2\Gyr\ (i.e. $z\lesssim 0.17$, \textit{red circles}), a period roughly corresponding to the dynamical time of the haloes in our sample. Typically, only $10-20~$per cent by mass of the X-ray luminous gas presently in galactic coronae \textit{ever} reached star forming densities. The fraction that attained those densities within a dynamical time of the present day - and could potentially therefore form part of an outflow at the present epoch - is smaller still, typically less than $2~$per cent. However, an interesting trend is seen as a function of halo mass, such that lower mass haloes do show a large scatter in this quantity. In a small number of cases, outflows dominate the overall mass of X-ray luminous gas. 

As shown in the right-hand panel, the \textit{luminosity} fraction contributed by these particles is disproportionately large for their mass. This is not unexpected since much of this gas remains relatively dense at the centre of the system even after being heated. Moreover, having formed part of the ISM, these particles are typically more metal rich, boosting their emissivity. Typically, however, gas that has \textit{recently} been heated out of the ISM and into the X-ray luminous phase contributes less than half of the overall soft X-ray luminosity of galactic coronae. 

\begin{figure*}
\includegraphics[width=\columnwidth]{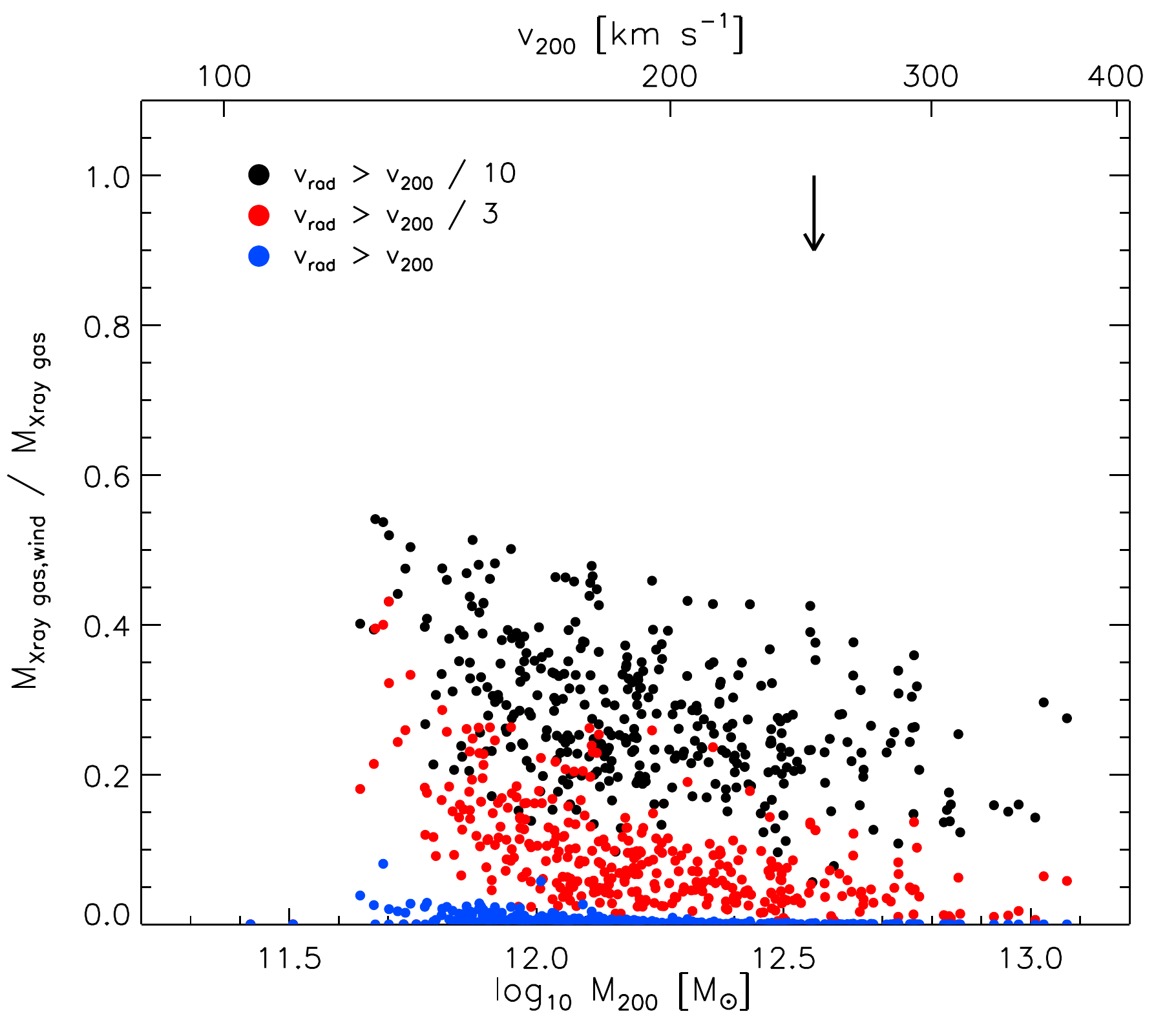}
\includegraphics[width=\columnwidth]{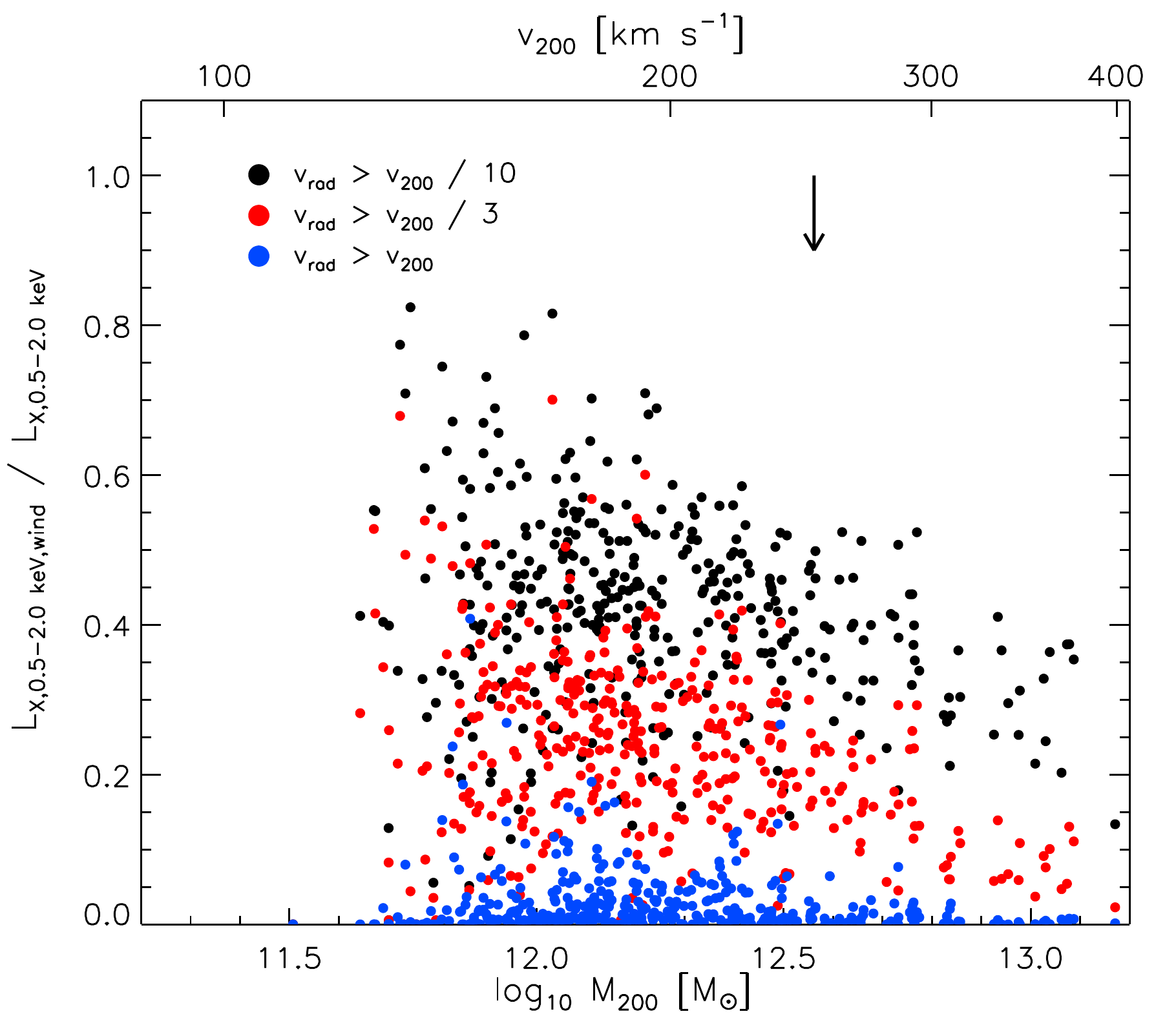}
\caption{The fraction of X-ray luminous mass (\textit{left-hand panel}) and soft X-ray luminosity (\textit{right-hand panel}) contributed by gas in our simulated galaxies that has a positive radial velocity at $z=0$. Each galaxy contributes three dots (one each of black, red and blue) to each plot. We delimit `slow' (\textit{black dots}), `intermediate' (\textit{red dots}) and `fast' (\textit{blue dots}) outflows with velocity cuts scaled to the system circular velocity, as shown in the legend. The down arrow at $v_{200}=250\kms$ marks the approximate scale above which supernova-driven winds become inefficient at regulating star formation (see C09). The plots are consistent with Fig.~\ref{fig:on_eos}: relatively little X-ray luminous mass is found in fast outflows. However, outflowing gas is typically more luminous (per unit mass) than hydrostatic or inflowing gas, owing to its centrally concentrated distribution and relatively high metallicity. The absence of a transition in behavour at $v_{\rm c}\simeq 250\kms$ indicates that the X-ray luminosity of the simulated galaxies is not in general driven by the present-day star formation rate.}
\label{fig:in_wind}
\end{figure*}

A more direct measure to establish the importance of present-day outflows to the overall emissivity of hot coronae is to identify hot gas particles with positive radial velocities, $v_{\rm rad}$, with respect to the galactic centre. In Fig.~\ref{fig:in_wind} we show the fractional contribution of these particles to i) the X-ray luminous gas mass of haloes (\textit{left-hand panel}) and ii) the soft X-ray luminosity of haloes. We show the contribution from particles in winds that exceed three velocity cuts, measured in terms of the halo circular velocity at the virial radius: $v_{200}/10$ (\textit{black dots}), $v_{200}/3$ (\textit{red dots}) and $v_{200}$ (\textit{blue dots}). Only a modest fraction of the hot gas mass exhibits an appreciable positive radial velocity, and very little approaches the circular velocity. It is therefore safe to conclude that all hot gas in the simulated systems remains gravitationally bound to the halo, since the escape velocity is typically a factor of a few greater than the circular velocity. As also concluded from inspection of Fig.~\ref{fig:on_eos}, most of the mass of X-ray luminous gas is inflowing (or near hydrostatic), albeit with a trend in halo mass such that outflows contribute more to the X-ray luminous mass fraction in less massive haloes. Again, in common with Fig.~\ref{fig:on_eos}, outflows contribute a disproportionate fraction of the overall X-ray luminosity because their density and metallicity are typically greater than for inflowing gas.

In \S~\ref{sec:comp_with_obs} we demonstrated that the \gimic\ simulations reproduce the observed correlation between the present-day star formation rate, $\dot{M}_\star$, and the soft X-ray luminosity, $L_{\rm X,0.5-2.0 keV}$ (see Fig.~\ref{fig:Lx_SFR}). This observed scaling has been interpreted by several authors as evidence that the formation of hot coronae is linked to supernova-driven feedback associated with star formation. However, we have just seen that typically less than half of the X-ray luminosity is contributed by recently heated, outflowing gas. Moreover, no transition in this trend is seen close to $v_{200}\sim 250\kms$ (denoted by a down arrow in Fig.~\ref{fig:in_wind}), the scale at which haloes exhibit a sharp transition in star formation efficiency (see Fig.~8 and associated discussion in C09). The absence of such a trend indicates that present-day star formation (and its associated supernova feedback) in general does not drive the X-ray luminosity.  

\begin{figure}
\includegraphics[width=\columnwidth]{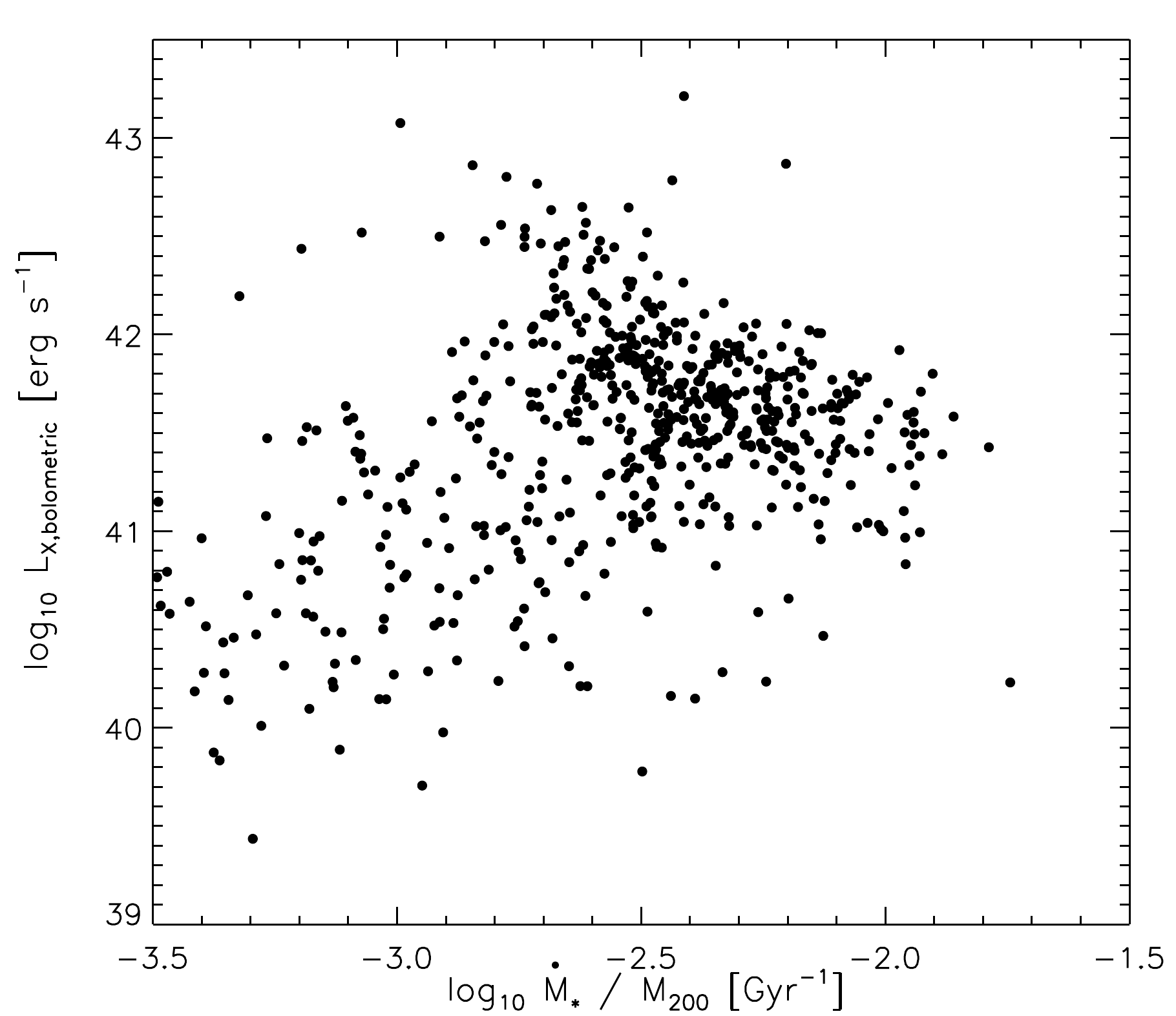}
\caption{The `bolometric' X-ray luminosity (0.02-50\keV) of hot gas coronae surrounding our simulated galaxies, as a function of their {\it specific} star formation rate (sSFR; $\dot{M}_{\star}/M_{200}$). Although we saw in Fig.~\ref{fig:Lx_SFR} that X-ray luminosity tracks the star formation rate, both quantities correlate with the system virial mass, $M_{200}$. With this scaling factored out, a weak correlation exhibiting a large scatter is evident for less massive systems, whilst for more massive systems the X-ray luminosity weakly anti-correlates with the sSFR. The $L_{\rm X}-\dot{M}_\star$ correlation is therefore driven primarily by the scaling of both quantities with $M_{200}$.} 
\label{fig:Lx_sSFR}
\end{figure}

The $\dot{M}_{\star}-L_{\rm X,0.5-2.0 keV}$ scaling has a simple, alternative explanation: both quantities scale with the virial mass of the halo, $M_{200}$. A direct illustration is provided by Fig.~\ref{fig:Lx_sSFR}, in which X-ray luminosity is plotted as a function of the halo \textit{specific} star formation rate of the galaxy, $\dot{M}_\star/M_{200}$. Note that, in this case, we plot the \textit{bolometric} X-ray luminosity\footnote{Whilst semantically inaccurate, the term `bolometric X-ray luminosity' is frequently used in the field. We adopt it here to mean the luminosity yielded by a plasma emission model, integrated over the 0.02-50\keV\ energy range.} as computed by \apec, to prevent a spurious drop in $L_{\rm X}$ for cases where a significant fraction of the emission falls outside of the soft X-ray waveband. With the halo mass dependence factored out of the star formation rate, it becomes clear that in the most luminous systems ($L_{\rm X,tot} \gtrsim 10^{41}\ergs$), the X-ray luminosity is weakly anti-correlated with $\dot{M}_\star/M_{200}$. For less luminous systems a modest positive correlation is visible, albeit with considerable scatter.

These trends are consistent with the conclusion we drew from Figs.~\ref{fig:on_eos} and \ref{fig:in_wind}, namely that in massive systems the X-ray luminosity is dominated by quasi-hydrostatic or inflowing gas, whilst in smaller counterparts outflowing gas contributes a significant, but not dominant, fraction of the luminosity. It is also noteworthy that observations are necessarily luminosity weighted.  As a result, X-ray data yield a rather unrepresentative view of the hot gas distribution, since the $\rho^2$ dependence of $L_{\rm X}$ leads to a strong biasing of the luminosity towards the halo centre, relative to the overall mass distribution of X-ray luminous gas. We find that, independently of halo mass, half of the X-ray luminosity is contained within $\sim 0.1-0.2 r_{200}$, whilst half of the X-ray luminous mass is contained within $\sim 0.6-0.7 r_{200}$. For this reason, the emission detected by X-ray telescopes is disproportionately weighted towards outflowing gas. Even with this, we predict that at least half of the X-ray luminosity detected around disc galaxies is from quasi-hydrostatic inflowing gas. High resolution spectroscopic X-ray observations enabled by future telescopes, such as \textit{NeXT/ASTRO-H} \citep{Takahashi_et_al_08_short} and eventually the \textit{International X-ray Observatory} (\textit{IXO}), have the potential to map directly the velocity structure of this gas, thus providing constraints on the fraction of hot gas that is entrained in outflows.

\subsubsection{What fuels star formation in discs: hot vs cold accretion?}
\label{sec:sfh_hot_cold}

In Fig.~\ref{fig:profiles} we showed that, at $z=0$, the cooling time of hot gas surrounding $L_\star$ disc galaxies is considerably longer than predicted by WF91. This deviation from the analytic model occurs because, in the hydrodynamic simualtions, star formation and supernova-driven feedback at $z\sim 1-3$ raise the adiabat of hot gas preventing its compression to high densities at later times. Since we find significant differences between the simulations and the canonical analytic picture of WF91, we are motivated to investigate whether the cooling of gas from a hot, quasi-hydrostatic reservoir is really what fuels star formation in galaxy discs at late times, as posited by WF91. As shown in Fig.~\ref{fig:Lx_SFR}, our simulated galaxies exhibit star formation rates that are consistent with observations; this indicates that the gas consumed by star formation in the simulations is efficiently replenished, or otherwise the cold gas in discs would be rapidly consumed.

It is less clear how this replenishment actually takes place. Recent studies employing hydrodynamic simulations have claimed that a significant fraction of a galaxy's stellar mass forms from gas that is channelled to the halo centre without ever experiencing a sustained accretion shock \citep{Keres_et_al_05,Dekel_et_al_09_short}. The thermalisation of the potential energy of this material occurs at significantly greater densities than those near the virial radius, drastically shortening its cooling time and suppressing the formation of a hot coronal phase. This scenario has spawned the term `cold accretion'. It corresponds roughly to the regime identified by WF91 in which the cooling time of the gas is much shorter than the dynamical time of the halo (see Fig.~2 of WF91) although WF91 considered only a spherically symmetric case whereas much of the `cold accretion' in the simulations seems to be associated with filaments.

It is instructive to assess, for our sample of simulated galaxies, the fraction of star formation from gas that has arrived at the disc i) having shocked and subsequently cooled (as posited by WF91 and similar models - the `hot mode') and ii) without having experienced a sustained accretion shock (`cold mode'). This is relatively straightforward to determine since our simulation code tags star particles with both their formation redshift and the maximum temperature attained by the gas particle from which they formed. We therefore adopt a simple definition of the two modes: stars that formed from gas whose temperature never rose above $T=2.0\times10^5\K$ (the minimum temperature we consider for X-ray emission) are considered to have formed via the cold mode; all others are considered to have formed via the hot mode. Note that this simple definition of `cold' and `hot' is not identical to that of e.g. \citet{Keres_et_al_05} and tends to exaggerate the amount of cold accretion since at least some of the gas below the temperature cut is likely to have been shock-heated to the virial temperature.

\begin{figure*}
\includegraphics[width=\columnwidth]{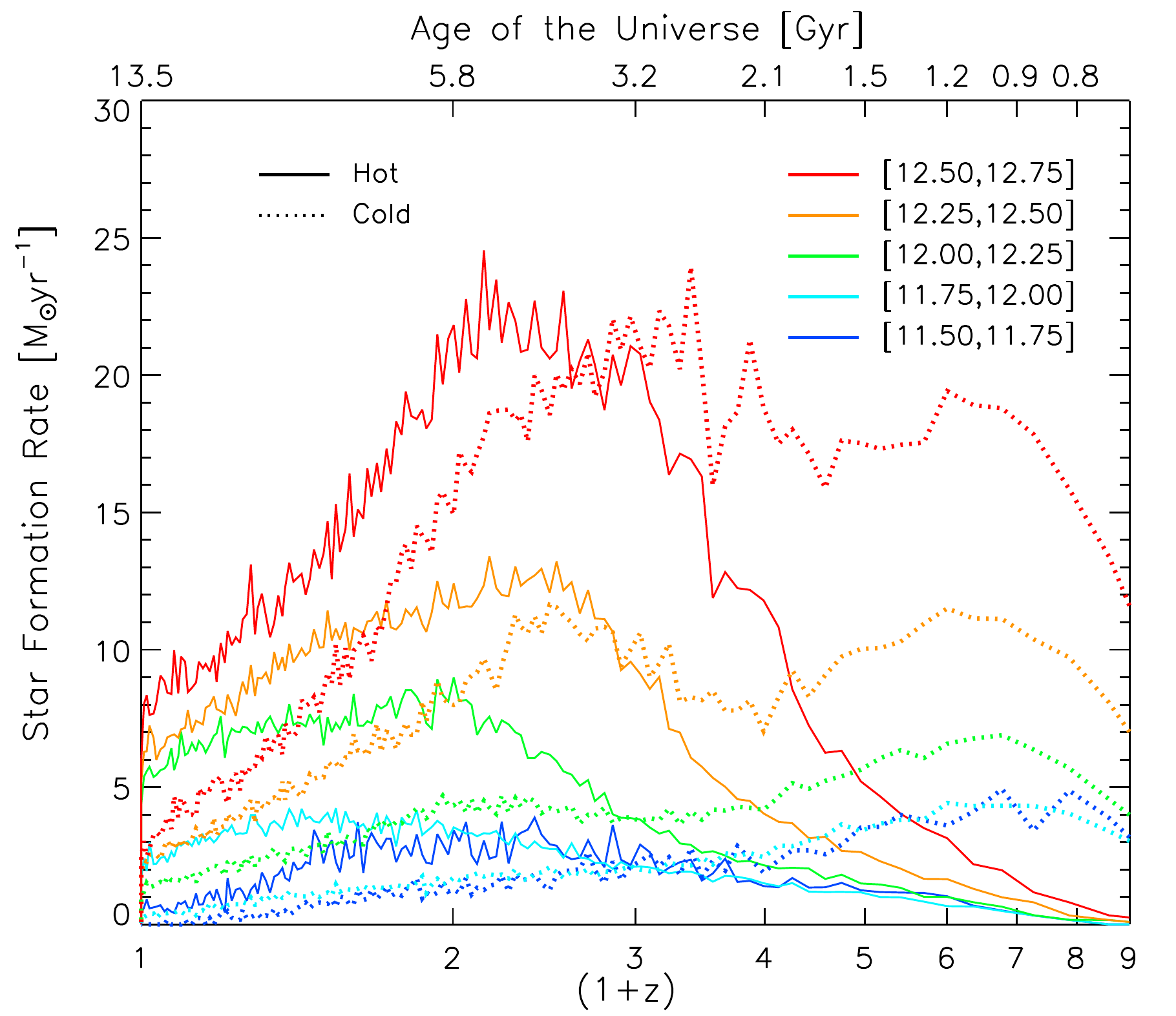}
\includegraphics[width=\columnwidth]{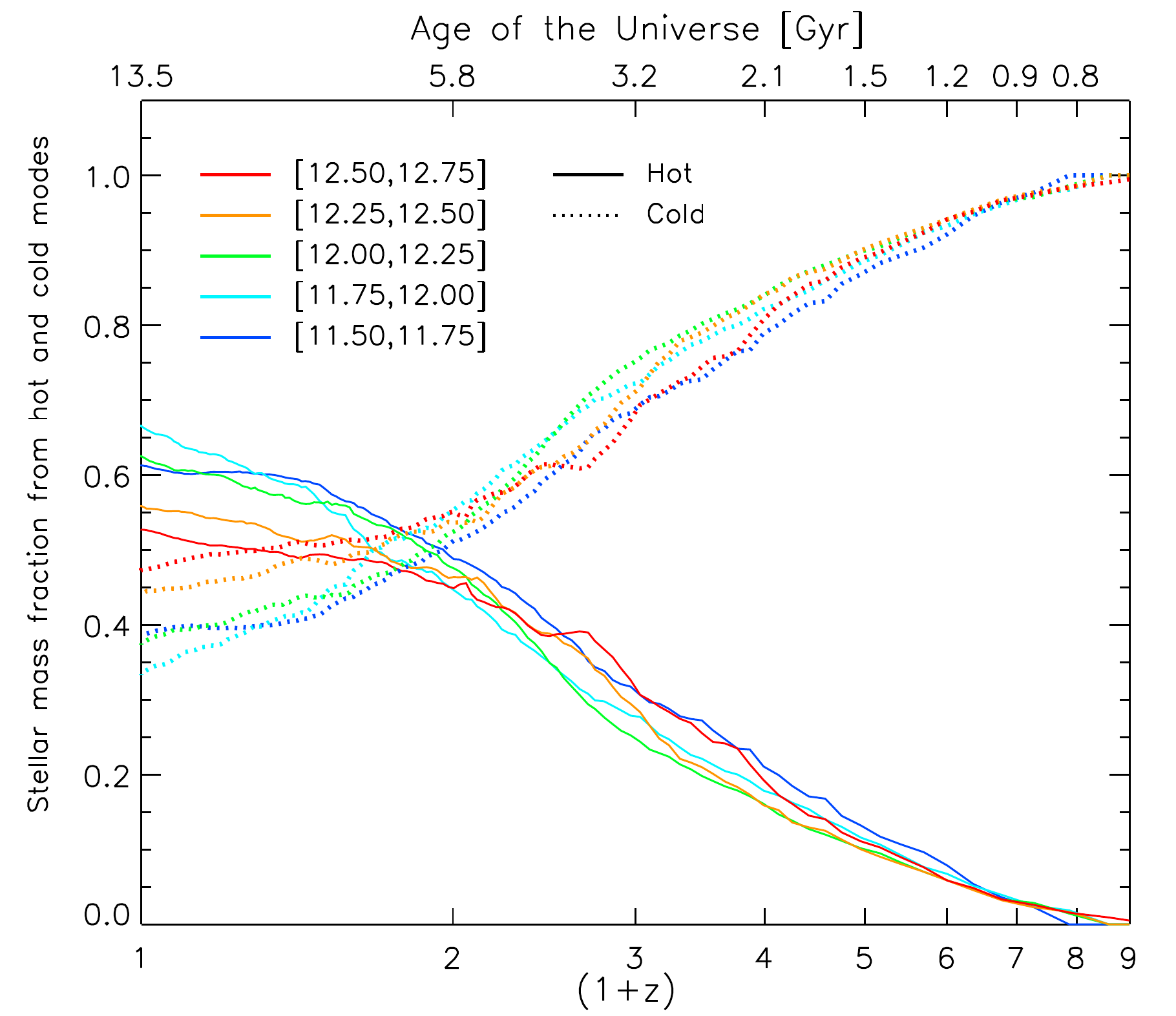}
\caption{The star formation history of all progenitors of our simulated galaxies, binned as in Figs.~\ref{fig:profiles} and \ref{fig:history}, decomposed into hot (\textit{solid lines}) and cold (\textit{dotted lines}) modes (see text for details). The left-hand panel shows the instantaneous star formation rate in each mode, whilst the right-hand panel shows the fraction of the total stellar mass in each bin that has formed in each mode by the given redshift. In every mass bin, star formation fuelled by cold mode accretion dominates at early epochs, whilst star formation from the hot mode becomes dominant at later epochs. The increase in hot mode star formation rate occurs sufficiently early that in every mass bin the stellar mass formed by the hot mode overtakes the cold mode mass at $z\simeq1$. Most stars today in our sample of $\sim L_\star$ galaxies formed from gas accreted in the hot mode.}
\label{fig:sfh_hot_cold}
\end{figure*}

The left-hand panel of Fig.~\ref{fig:sfh_hot_cold} compares the instantaneous star formation rate fuelled by the hot (\textit{solid lines}) and cold (\textit{dotted lines}) modes, as a function of redshift for all progenitors of our simulated galaxies. The galaxies are binned according to their virial mass at $z=0$ and their median star formation histories are plotted. The cold mode dominates early star formation in all mass bins, but in all cases the hot mode overtakes the cold mode during the interval $1<z<2$ (which corresponds roughly to half the age of the Universe). The right-hand panel of Fig.~\ref{fig:sfh_hot_cold} shows the stellar mass fraction formed from gas accreted in each mode up to a given redshift. The fraction of stars formed in each mode is almost independent of halo mass, in the range considered here. In all cases, most of the stars present in these haloes today formed from gas accreted in the hot mode. The fraction ranges from $\sim 65\%$ for the least massive haloes plotted in the figure to $\sim 55\%$ for the most massive. Weighting by the relative numbers of haloes of different mass, the fraction of stars formed from hot mode accretion in haloes with mass above the minimum shown in the figure ($M_{200}=10^{11.5}\Msun$) is $\sim 54\%$. The values of these quantities no doubt depend, to some extent, on the particular feedback scheme adopted in the simulations, but any scheme that gives a reasonable galaxy stellar mass function is likely to produce similar numbers. The dependence of various modes of star formation on the feedback model is being investigated in the \owls\ simulations by van de Voort et al. (in preparation). Beyond this, it will be important to check that the results are independent of the numerical hydrodynamics technique employed, whether it be SPH, AMR or a moving mesh scheme.

The dominance of hot mode star formation at low redshift is of particular relevance here because it supports the simple assumption of the WF91 model that disc formation at the present epoch is fuelled by the cooling and condensation of gas from a hot halo. This is in spite of the fact that, as we have shown here, the gas in the hot haloes lies on a higher adiabat than assumed by WF91 (i.e. it is hotter, less dense, and has a longer cooling time). Note that this general picture is not inconsistent with the observation that extra-planar high-velocity clouds (HVCs) are abundant in the Milky Way and nearby disc galaxies and may be transporting a significant mass of cold gas to their discs \citep[e.g.][]{Oort_70,Kerp_et_al_96,Wakker_and_van_Woerden_97,Wakker_et_al_99_short,Tripp_et_al_03_short,Thiliker_et_al_04,Miller_et_al_09}.  Such clouds may well have formed from thermal instability and fragmentation of hot coronal gas \citep[e.g.][]{Maller_and_Bullock_04,Booth_and_Theuns_07}.

\subsection{Comparison with other theoretical work}
\label{sec:comparison_other_theory}

At this juncture it is instructive to draw comparisons with previous attempts to compute the X-ray luminosity of hot coronae of disc galaxies at $z=0$. The simulations discussed in this context by \citet{Rasmussen_et_al_09}, initially presented by \citet{Sommer-Larsen_Romeo_and_Portinari_05} and \citet{Romeo_et_al_06}, and which were an updated version of those discussed in a similar context by \citet{Toft_et_al_02}, arrived at seemingly similar conclusions to those presented here.

However, the similarity between the two sets of conclusions is probably coincidental because of the very different treatments of baryonic physics in the two studies. In particular, the supernova feedback scheme implemented by \citet{Rasmussen_et_al_09} was only effective at driving galactic scale outflows at $z\gtrsim 4-5$. As a result, star formation in their galaxies was not self-regulating and ended up with stellar mass fractions ($f_\star \equiv M_\star/M_{200}$) roughly twice as large as those in our \gimic\ sample (J. Sommer-Larsen, private communication). Such high stellar fractions are ruled out by analyses of the Sloan survey \citep[e.g.][]{Guo_et_al_09}, indicating that the X-ray luminosities produced by galaxies in the simulations of \citet{Toft_et_al_02} and \citet{Rasmussen_et_al_09} are lower than predicted by WF91 primarily because of overcooling and excessive star formation. According to our results, these studies underestimate the role of feedback, which we found to be dominant, in explaining why the X-ray luminosities of disc galaxies are much smaller than predicted by WF91.

In principle, the effects of radiative cooling and feedback can be included in semi-analytic models, as shown by \citet{Bower_McCarthy_and_Benson_08}, who attempted to construct a self-consistent model of the heating of the intracluster medium by AGN and its effect on the subsequent cooling of the gas. This study focussed on the X-ray properties of groups and clusters, but the methodology is applicable to galaxies as well.

\section{Summary and Discussion}
\label{sec:summary}

The existence of hot, X-ray luminous gas reservoirs surrounding present day Milky Way-like galaxies is a key prediction of the galaxy formation model developed by \citet[][WF91]{White_and_Frenk_91}. This prediction is important because the WF91 model, and the many analytic and semi-analytic models based on elements of it, successfully reproduce a broad range of observed galaxy properties. However, evidence in support of this particular prediction has, in general, not been forthcoming. In this paper, we have used the \textit{Galaxies-Intergalactic Medium Interaction Calculation} \citep[][C09]{Crain_et_al_09_short} to investigate this challenge to the standard picture of galaxy formation using altogether different techniques. About half the galaxies that form in these five simulations are disc-dominated (an interesting fact in its own right which we will explore in detail in a further paper). Of these, 458 have stellar masses greater than $10^{10}\hMsun$ and make up the sample we have investigated here.

\citet[][B00]{Benson_et_al_00} searched for the predicted extended emission around three promising candidate disc galaxies using the \rosat\ telescope, but they failed to detect any.  Subsequent extensive surveys with the \xmm\ and \chandra\ telescopes finally succeeded in detecting diffuse emission in the soft (0.5-2.0\keV) X-ray band around some disc galaxies, but at much lower levels than predicted by WF91. The larger number of non-detections in these surveys has tightly constrained upper limits on this kind of emission. The observational data now offer a clearer picture than was available to B00 who were the first to highlight the potential implications of the absence of X-ray emission for galaxy formation theory. In sum, the data indicate that the soft X-ray luminosities of disc galaxies are one to two orders of magnitude lower than predicted and that they broadly correlate with both the $K$-band luminosity and the star formation rate of the galaxy.

The \gimic\ hydrodynamic simulations make different assumptions and approximations to those of the WF91 model. However, we find that these simulations also generate X-ray coronae around disc galaxies, although these are much weaker that predicted by WF91 for the reasons summarized below. Many of the bulk X-ray properties of the simulated galaxies agree with observations. In particular, the 458 galaxies in our sample exhibit similar scaling relations to the data. For this comparison, we used a compilation of X-ray measurements and upper limits from several studies \citep{Benson_et_al_00,Strickland_et_al_04,Wang_05,Tullmann_et_al_06,Sun_et_al_07,Owen_and_Warwick_09,Rasmussen_et_al_09} and extracted $K$-band luminosites from the \twomass\ database and disc rotation velocities (derived from inclination-corrected 21-cm measurements) from the \textsc{HyperLeda} database. We find that the galaxies in the simulations broadly reproduce the observed scaling and scatter of the $L_{\rm X}-L_{\rm K}$, $L_{\rm X}-v_{\rm rot}$ and $L_{\rm X}-\dot{M}_\star$ relations at $z=0$.

There are two main reasons why the X-ray emission from the \gimic\ galaxies is much weaker that predicted by WF91. Firstly, the mass of hot gas within the galaxies' haloes is, typically, lower than in the WF91 model. This is partly because some of the gas has been consumed into stars and partly because some gas has been blown out in winds. Secondly, and most importantly, the distribution of the hot gas is much less concentrated than the distribution of dark matter, violating one of the main assumptions of the WF91 model. The gas is more extended than expected because its entropy has been significantly raised. For galaxies like the Milky Way and fainter, the increase in entropy is mostly due to the injection of energy generated by supernovae. The gas is placed in a high adiabat at early times, $z\sim 1-3$, when the star formation peaks.

The entropy of the corona can also be affected by radiative cooling. Cooling acts to reduce the gas temperature by selectively removing the lowest entropy gas and locking it into stars, thus raising the mean entropy of the remaining coronal gas \citep[e.g.][]{Voit_et_al_02}. The effect of radiative cooling in this respect is therefore not dissimilar to that of outflows. However, we have shown that the reduction of the hot gas density by radiative cooling is only significant for the most massive galaxies in our sample. In this regime, however, our simulations overestimate the effects of radiative cooling because they do not include additional energetic feedback processes such as AGN. This shortcoming is manifest, for example, in the high stellar fractions of the largest galaxies which are about a factor of 2 larger than inferred observationally \citep{Guo_et_al_09}. By contrast, the median stellar fractions of the Milky Way-like galaxies and smaller in our simulations are in very good agreement with these data, although the scatter in the simulations appears perhaps to be too large. Simulations that include the effects of AGN are required to understand the X-ray properties of the gas associated with the most massive galaxies.

The main difference between the hydrodynamic simulations and the semi-analytic models based on the WF91 formalism is that the simulations follow the kinematics and dynamics of the gas self-consistently. Although outflows are often included in semi-analytic models \citep[see e.g.][]{Kauffmann_99,Cole_et_al_00,Bower_et_al_01,Baugh_et_al_05, Bower_et_al_06,Croton_et_al_06_short,Bertone_DeLucia_and_Thomas_07}, the readjustment of the gas to energy injection is not generally taken into account. One exception is the recent work by \citet{Bower_McCarthy_and_Benson_08} which attempts to follow self-consistently the heating of the intracluster medium by AGN and its effects on the subsequent cooling of the gas. The key contribution of the full hydrodynamic treatment is the ability to follow the way in which the entire structure of the gas corona adjusts in order to accommodate the increase in entropy associated with the energy injection by supernovae (or AGN). It is important to note, however, that the \gimic\ simulations assume a specific model for energy feedback. Although this appears well motivated, it is essential to investigate how the X-ray properties of the gas depend on the implementation of energy feedback processes. Whilst methods such as that developed by \citet{Bower_McCarthy_and_Benson_08} provide a blueprint for starting to explore the reaction of gas to energy injection in a computationally inexpensive fashion, we note that model-independent conclusions can only be arrived at with suites of hydrodynamical simulations featuring a wide range of possible feedback treatments (an approach adopted by the \owls\ simulations, albeit at lower resolution than required here). We intend to pursue this methodology with high-resolution resimulations of individual galaxies in future studies.

In spite of the approximations inherent in the analytic treatment of gas dynamics in the WF91 model, key results of this model are borne out by the \gimic\ simulations. The predicted quasi-hydrostatic hot coronae around $L^\star$ disc galaxies do indeed form, as anticipated, by the shock heating of infalling gas as the dark matter halo develops. In all but a few low mass galaxies, emission from these coronae dominates the galactic diffuse X-ray luminosity.

Galactic winds contribute to the X-ray luminosity but, in general, their emission is subdominant. Winds represent a small fraction of the coronal gas, but since they tend to be hotter, denser and more metal-rich than the quasi-hydrostatic component, they contribute disproportionally to the X-ray luminosity. Since most X-ray observations target the central parts of disc galaxies, where the wind gas is concentrated, and since the observed X-ray luminosity correlates well with the star formation rate, it is easy to misinterpret the importance of winds to the overall X-ray emission. In fact, in our simulations the X-ray luminosity also correlates with the star formation rate, just as in the data, even though the emission is dominated by the quasi-hydrostatic corona. This correlation occurs because both the star formation rate and the total amount of hot gas depend on the mass of the halo. The simulations do produce a few examples of bright galaxies in which outflowing gas contributes the bulk of the X-ray luminosity. These may be counterparts of starburst galaxies such as M82 \citep[whose intense star formation seems to be associated with interaction with M81, see e.g.][]{Chynoweth_et_al_08}. In our simulations, star formation in $L^\star$ disc galaxies today is mostly fuelled, as found by WF91, by the cooling of gas out of the hot corona. This can occur either through a smooth cooling flow \citep[e.g.][]{Thomas_et_al_86} or through fragmentation into high-velocity clouds \citep[e.g.][]{Maller_and_Bullock_04,Booth_and_Theuns_07}.

Our simulations indicate that hot outflows do not, in general, dominate the overall soft X-ray luminosity of isolated disc galaxies. We therefore expect gravitational inflow to dominate over supernova-driven outflows, and the majority of a galaxy's soft X-ray luminosity to stem from hot, quasi-hydrostatic gas at low surface brightnesses. This physical picture is, in principle, testable with future X-ray observatories with instrumentation of high sensitivity and spectral resolution, such as \textit{NeXT/ASTRO-H} and \textit{IXO}, because coronal gas dynamics follow directly from X-ray line diagnostics. We therefore expect these facilities directly to verify, or rule out, the key r\^ole in the formation of disc galaxies that theoretical models assign to hot gaseous coronae.

\section*{Acknowledgements}
\label{sec:acknowledgements}

We thank Richard Bower, Claudio Dalla Vecchia, Cedric Lacey, Trevor Ponman and Simon White for insightful discussions, the anonymous referee for many useful suggestions that improved the paper, and Jesper Rasmussen and Jesper Sommer-Larsen for providing details of their simulations. RAC acknowledges the hospitality of the Institute for Computational Cosmology, Durham, and the Institute of Astromomy, Cambridge, where parts of this work were conducted. RAC is supported by the Australian Research Council via a Discovery Project grant. IGM acknowledges support from a Kavli Institute Fellowship at the University of Cambridge. CSF acknowledges a Royal Society Research Merit award. This work was supported in part by an STFC rolling grant to the ICC. The simulations presented here were carried out using the HPCx facility at the Edinburgh Parallel Computing Centre (EPCC) as part of the EC's DEISA `Extreme Computing Initiative', and the Cosmology Machine at the Institute for Computational Cosmology of Durham University. This study makes use of the \textsc{HyperLeda} database (http://leda.univ-lyon1.fr), and data products from the Two Micron All Sky Survey, which is a joint project of the University of Massachusetts and \ipac/Caltech, funded by NASA and the NSF.


\bsp

\begin{appendix}
\medskip
\section{Convergence}

\begin{figure}
\includegraphics[width=\columnwidth]{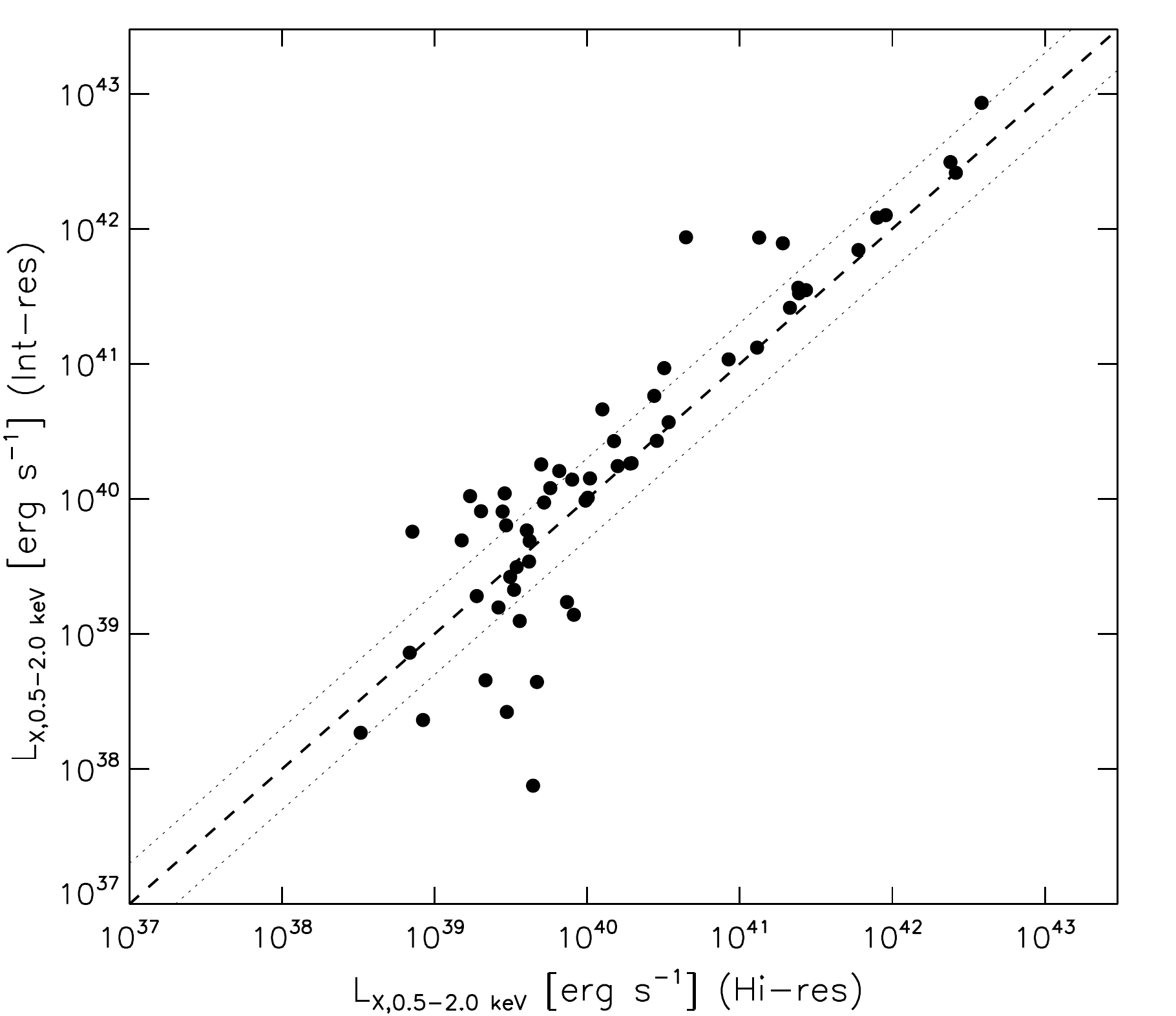}
\caption{The correspondence between the soft X-ray luminosity ($L_{\rm X,0.5-2.0~keV}$) of galaxies in the intermediate-resolution $-2\sigma$ \gimic\ simulation, and that of their counterparts in the high-resolution realisation of the same region. The dashed line denotes the 1:1 relation, and the dotted lines are offset higher and lower by a factor of 2.}
\label{fig:convergence}
\end{figure}

Our sample of simulated galaxies is drawn from the five intermediate-resolution \gimic\ simulations, which form a complete set at $z=0$. In addition, the \gimic\ suite includes one simulation to $z=0$, that of the low density $-2\sigma$ region, with 8 times better resolution. This simulation allows an assessment of the numerical convergence of our results. In C09 we used this and other high-resolution simulations at high redshift to show that the star formation rate density in the intermediate-resolution simulations has converged for $z \lesssim 6$, and that the halo specific star formation rate at $z=0$, $\dot{M}_\star/M_{200}$, has converged for haloes with $v_{200} \gtrsim 100\kms$.

The main quantity of interest here is the soft X-ray luminosity of galaxies at $z=0$. For each galaxy in our sample drawn from the $-2\sigma$ region at intermediate resolution, we have identified its counterpart in the high-resolution realisation. We assess convergence by comparing the soft X-ray luminosity of the objects at intermediate and high resolution in Fig.~\ref{fig:convergence}. The dashed line traces the locus of $L_{\rm X}^{\rm int-res} = L_{\rm X}^{\rm hi-res}$, whilst the upper and lower dotted lines are offset higher and lower by a factor of 2.

The data exhibit excellent convergence properties, with more than 50~percent of systems having soft X-ray luminosities in the intermediate- and high-resolution runs that agree to within a factor of 2, and 85~percent that agree to within a factor of 5, over 4 decades in $L_{\rm X}$. The degree of convergence in $L_{\rm X}$ is more than adequate for the purposes of this work. For example, the spread in $L_{\rm X}$ for galaxies of a given $L_{\rm K}$ in the observational data is typically much greater than the typical difference in $L_{\rm X}$ for a given galaxy simulated at intermediate- and high-resolution. Thus, we expect that our main conclusions would remain unchanged if they had been obtained from a full set of 5 \gimic\ regions at high resolution.

\end{appendix}
\label{lastpage}
\end{document}